\documentclass[ALICE,manyauthors]{cernphprep}
\usepackage[comma,square,numbers,sort&compress]{natbib}

\usepackage{upgreek}
\usepackage{lineno}
\usepackage{xspace}
\usepackage{xcolor}
\usepackage{amsmath} 
\usepackage{tabularx}
\usepackage[colorlinks=true, linkcolor=blue, citecolor=blue, urlcolor=blue]{hyperref}
\usepackage[T1]{fontenc}
\usepackage[utf8]{inputenc}
\usepackage{orcidlink} 

\begin{document}
%

\newcommand{\pp}           {pp\xspace}
\newcommand{\ppbar}        {\mbox{$\mathrm {p\overline{p}}$}\xspace}
\newcommand{\XeXe}         {\mbox{Xe--Xe}\xspace}
\newcommand{\PbPb}         {\mbox{Pb--Pb}\xspace}
\newcommand{\pA}           {\mbox{pA}\xspace}
\newcommand{\pPb}          {\mbox{p--Pb}\xspace}
\newcommand{\AuAu}         {\mbox{Au--Au}\xspace}
\newcommand{\dAu}          {\mbox{d--Au}\xspace}

\newcommand{\s}            {\ensuremath{\sqrt{s}}\xspace}
\newcommand{\snn}          {\ensuremath{\sqrt{s_{\mathrm{NN}}}}\xspace}
\newcommand{\pt}           {\ensuremath{p_{\rm T}}\xspace}
\newcommand{\meanpt}       {$\langle p_{\mathrm{T}}\rangle$\xspace}
\newcommand{\ycms}         {\ensuremath{y_{\rm CMS}}\xspace}
\newcommand{\ylab}         {\ensuremath{y_{\rm lab}}\xspace}
\newcommand{\etarange}[1]  {\mbox{$\left | \eta \right |~<~#1$}}
\newcommand{\yrange}[1]    {\mbox{$\left | y \right |~<~#1$}}
\newcommand{\dndy}         {\ensuremath{\mathrm{d}N_\mathrm{ch}/\mathrm{d}y}\xspace}
\newcommand{\dndeta}       {\ensuremath{\mathrm{d}N_\mathrm{ch}/\mathrm{d}\eta}\xspace}
\newcommand{\avdndeta}     {\ensuremath{\langle\dndeta\rangle}\xspace}
\newcommand{\dNdy}         {\ensuremath{\mathrm{d}N_\mathrm{ch}/\mathrm{d}y}\xspace}
\newcommand{\Npart}        {\ensuremath{N_\mathrm{part}}\xspace}
\newcommand{\Ncoll}        {\ensuremath{N_\mathrm{coll}}\xspace}
\newcommand{\dEdx}         {\ensuremath{\textrm{d}E/\textrm{d}x}\xspace}
\newcommand{\RpPb}         {\ensuremath{R_{\rm pPb}}\xspace}

\newcommand{\nineH}        {$\sqrt{s}~=~0.9$~Te\kern-.1emV\xspace}
\newcommand{\seven}        {$\sqrt{s}~=~7$~Te\kern-.1emV\xspace}
\newcommand{\twoH}         {$\sqrt{s}~=~0.2$~Te\kern-.1emV\xspace}
\newcommand{\twosevensix}  {$\sqrt{s}~=~2.76$~Te\kern-.1emV\xspace}
\newcommand{\five}         {$\sqrt{s}~=~5.02$~Te\kern-.1emV\xspace}
\newcommand{\twosevensixnn}{$\sqrt{s_{\mathrm{NN}}}~=~2.76$~Te\kern-.1emV\xspace}
\newcommand{\fivenn}       {$\sqrt{s_{\mathrm{NN}}}~=~5.02$~Te\kern-.1emV\xspace}
\newcommand{\LT}           {L{\'e}vy-Tsallis\xspace}
\newcommand{\GeVc}         {Ge\kern-.1emV/$c$\xspace}
\newcommand{\MeVc}         {Me\kern-.1emV/$c$\xspace}
\newcommand{\TeV}          {Te\kern-.1emV\xspace}
\newcommand{\GeV}          {Ge\kern-.1emV\xspace}
\newcommand{\MeV}          {Me\kern-.1emV\xspace}
\newcommand{\GeVmass}      {Ge\kern-.2emV/$c^2$\xspace}
\newcommand{\MeVmass}      {Me\kern-.2emV/$c^2$\xspace}
\newcommand{\lumi}         {\ensuremath{\mathcal{L}}\xspace}

\newcommand{\ITS}          {\rm{ITS}\xspace}
\newcommand{\TOF}          {\rm{TOF}\xspace}
\newcommand{\ZDC}          {\rm{ZDC}\xspace}
\newcommand{\ZDCs}         {\rm{ZDCs}\xspace}
\newcommand{\ZNA}          {\rm{ZNA}\xspace}
\newcommand{\ZNC}          {\rm{ZNC}\xspace}
\newcommand{\SPD}          {\rm{SPD}\xspace}
\newcommand{\SDD}          {\rm{SDD}\xspace}
\newcommand{\SSD}          {\rm{SSD}\xspace}
\newcommand{\TPC}          {\rm{TPC}\xspace}
\newcommand{\TRD}          {\rm{TRD}\xspace}
\newcommand{\VZERO}        {\rm{V0}\xspace}
\newcommand{\VZEROA}       {\rm{V0A}\xspace}
\newcommand{\VZEROC}       {\rm{V0C}\xspace}
\newcommand{\Vdecay} 	   {\ensuremath{V^{0}}\xspace}

\newcommand{\ee}           {\ensuremath{e^{+}e^{-}}} 
\newcommand{\pip}          {\ensuremath{\pi^{+}}\xspace}
\newcommand{\pim}          {\ensuremath{\pi^{-}}\xspace}
\newcommand{\kap}          {\ensuremath{\rm{K}^{+}}\xspace}
\newcommand{\kam}          {\ensuremath{\rm{K}^{-}}\xspace}
\newcommand{\pbar}         {\ensuremath{\rm\overline{p}}\xspace}
\newcommand{\kzero}        {\ensuremath{{\rm K}^{0}_{\rm{S}}}\xspace}
\newcommand{\lmb}          {\ensuremath{\Lambda}\xspace}
\newcommand{\almb}         {\ensuremath{\overline{\Lambda}}\xspace}
\newcommand{\Om}           {\ensuremath{\Omega^-}\xspace}
\newcommand{\Mo}           {\ensuremath{\overline{\Omega}^+}\xspace}
\newcommand{\X}            {\ensuremath{\Xi^-}\xspace}
\newcommand{\Ix}           {\ensuremath{\overline{\Xi}^+}\xspace}
\newcommand{\Xis}          {\ensuremath{\Xi^{\pm}}\xspace}
\newcommand{\Oms}          {\ensuremath{\Omega^{\pm}}\xspace}
\newcommand{\degree}       {\ensuremath{^{\rm o}}\xspace}

\newcommand{\pT}{\ensuremath{p_{\mathrm{T}}}\xspace}
\newcommand{\Lam}{\ensuremath{\Lambda}\xspace}

\newcommand{\ppL}{p--p--\ensuremath{\Lambda}\xspace}
\newcommand{\Q}{\ensuremath{Q_{\mathrm{3}}}\xspace}

\begin{titlepage}
\PHyear{2026}       
\PHnumber{237}      
\PHdate{04 August}  

\title{Three-baryon femtoscopy as an effective 3$\rightarrow$3 scattering experiment}
\ShortTitle{Three-Baryon Femtoscopy}   

\Collaboration{ALICE Collaboration\thanks{See Appendix~\ref{app:collab} for the list of collaboration members}}
\ShortAuthor{ALICE Collaboration} 

\begin{abstract}

Scattering experiments have long been the gold standard for constraining hadron--hadron interactions, providing direct information on the angular momentum and spin dependence over a wide range of kinematic configurations. However, experimental constraints on three-body dynamics remain limited, specifically for unbound systems and systems involving short-lived hadrons. 
In this work, the three-proton correlation function is measured in pp collisions at $\sqrt{s}=13.6$ TeV with ALICE at the LHC and presented as a novel approach to access hadronic interactions in three-body systems. A new analysis strategy is employed to isolate the p--p--p contribution to the correlation function by correcting for background channels and experimental effects, and enabling a direct comparison with state-of-the-art three-body continuum calculations. 
The extracted correlation function provides the first direct access to the isospin $3/2$ three-body system. The measured observable is found to be sensitive to the partial-wave structure of the nucleon--nucleon interaction and indicates that the nuclear interaction acts even at high angular momentum and parity states of the three-body system, revealing an effective long-range attractive component, observed experimentally for the first time in a three-proton continuum system. Hence, three-hadron femtoscopy emerges as an effective 3$\rightarrow$3 scattering experiment with three unbound hadrons in initial and final states. The copious production of hyperons at the modern high-energy colliders ensures the possibility of extending such measurements beyond nucleons, opening a new avenue for future precision studies of three-body dynamics in the strangeness sector.

\end{abstract}
\end{titlepage}

\setcounter{page}{2} 

\section{Introduction} 
The strong interaction is the dominant force responsible for the structure of all nuclear systems in the Universe, from the lightest nuclei to the dense matter in neutron star interiors. Over the years, the development of sophisticated interaction models within effective field theories (EFT) has demanded increasingly precise experimental data to benchmark the growing level of theoretical details necessary to describe the complex low-energy behavior of the nuclear interaction~\cite{Hammer:2019poc,Epelbaum:2002vt,Hergert:2020bxy,Epelbaum:2008ga}. Differential cross sections measured over a broad energy range have been crucial to constraining the angular-momentum dependence of the nucleon--nucleon (NN) interaction~\cite{Epelbaum:2008ga,Stoks:1993tb,Stoks:1994wp}. Moreover, precision measurements of the binding energies of light nuclei presented a discrepancy with calculations based only on pairwise interactions, revealing a substantial effect of the order of 10\% from three-nucleon forces~\cite{LENPIC:2018ewt,Piarulli:2017dwd}. The necessity of a three-body component was later confirmed in the studies of proton--deuteron (p--d) scattering~\cite{Gloeckle:1995jg,Kievsky:2001fq,Kievsky:2003eu,Witala:2022rzl,Girlanda:2023znc}.

The same level of description has not yet been achieved for other strongly interacting particles. In particular, the studies of hyperon interactions with nucleons are important for understanding how their expected appearance in dense matter, such as in the core of neutron stars, could modify the corresponding equation of state~\cite{Logoteta:2019utx,Vidana:2024ngv}. However, low-energy scattering experiments with hyperons are particularly difficult to carry out due to their short lifetimes ($\tau < 10$ cm/$c$~\cite{ParticleDataGroup:2024cfk})~\cite{Eisele:1971mk,Alexander:1968acu,Sechi-Zorn:1968mao,BESIII:2024geh,CLAS:2021gur,J-PARCE40:2022nvq} and bound systems of $\Lambda$~\cite{Gal:2016boi}, $\Sigma$~\cite{Nagae:1998tj}, and $\Xi^- $~\cite{Nakazawa:2015joa,J-PARCE07:2020xbm,Yoshimoto:2021ljs} hyperons with nucleons provide limited constraints to theory, particularly for three-body forces~\cite{Gal:2016boi,Haidenbauer:2019boi}. Dedicated experimental programs are currently ongoing~\cite{ALICE:2022wwr,Aberle:2931010}, and new experimental approaches are highly desirable.

In recent years, femtoscopy has been established as a complementary approach to study the strong interaction between hadrons by measuring the momentum correlation of particles emitted at separations of a few femtometers at collider facilities~\cite{Fabbietti:2020bfg,ALICE:2020ibs,HADES:2010izx,STAR:2015kha}. This method overcomes the limitations of traditional scattering experiments, as strange hadrons can be produced abundantly in high-energy collisions and interact with other particles before decaying. One remarkable example is the p--$\Lambda$ correlation measurement performed by the ALICE Collaboration in pp collisions at $\sqrt{s}=13$ TeV at the Large Hadron Collider (LHC)~\cite{ALICE:2021njx}, which provided stringent constraints on the p--$\Lambda$ interaction, complementing existing scattering data~\cite{Mihaylov:2023ahn}. The p--d correlations have also been studied in the same collision system, demonstrating that the three-body dynamics can be resolved~\cite{Viviani:2023kxw, ALICE:2023bny}. Since the deuteron is a bound system of a neutron and a proton in an isospin-singlet and spin-triplet state, only these quantum number configurations are accessible for this system. Measuring three-particle correlations instead would not only extend the number of measurable configurations but could also provide the first experimental constraints on the unbound (continuum) energy states of the three baryons, and validate calculations used to infer the properties of possible quasi-bound states with three and four neutrons~\cite{Higgins:2020pbe,Higgins:2020avy}. 
The feasibility of such measurements has already been proven by the ALICE Collaboration using data recorded during the LHC Run 2 data-taking period (2015--2018), but the available data were statistically limited, and theoretical predictions were not available~\cite{ALICE:2022boj,ALICE:2023gxp,ALICE:2025aur,ALICE:2022wpn}. Recently, a numerical framework to calculate three-body correlation functions has been developed, describing the measured process as elastic three-unbound baryon scattering or so-called three-to-three (3$\rightarrow$3) reactions~\cite{Kievsky:2023maf,Garrido:2024pwi,Garrido:2025lar}. Moreover, theoretical studies of the \ppL system have shown that information on the three-body interaction can also be imprinted in the correlation signal~\cite{Garrido:2024pwi}. Thus, the full theoretical and experimental framework of three-body femtoscopy would establish a completely new class of continuum observables of three-hadron dynamics, including hyperons, beyond the reach of existing experimental approaches.

In this work, the three-proton (p--p--p) correlation function is measured in pp collisions at the center-of-mass energy of $\sqrt{s}=13.6$ TeV recorded by the ALICE Collaboration during the LHC Run 3 data-taking period. Proving the sensitivity of correlation functions to the partial-wave structure of the nuclear force would establish these measurements as an alternative to scattering data and would give access for the first time to the three-body continuum dynamics. Since the NN interactions are known with high precision, a three-nucleon system serves as a controlled benchmark to quantify the sensitivity of the correlation observable to the inclusion of the two- and three-body angular-momentum components in the calculations. The developed analysis strategy and the precision of the measured correlation function enable the extraction of the genuine p--p--p correlation function unhindered by background channels and experimental effects, and thus permit a direct comparison to theoretical predictions for the first time. The full-fledged three-body calculations show that the presented measurement captures the angular-momentum dependence of the interaction and indeed acts as an effective 3$\rightarrow$3 scattering experiment. Moreover, an effective long-range attractive component is observed experimentally for the first time in a three-proton continuum system. This work opens the avenue for new programs of precision studies of any three-hadron interactions that can be pursued by the current and future collider facilities.

\section{Experimental method and analysis}\label{sec:analysis}
Hadrons produced in high-energy collisions may undergo final-state interactions when they are emitted at femtometer-scale separations, resulting in momentum correlations among them. Such correlations can be studied using the femtoscopy method, in which the correlation function is the key observable. For two-hadron systems, it is defined as $C(k^*)=\mathcal{N}\ N_{\mathrm{same}}(k^*)/N_{\mathrm{mixed}}(k^*)$, where $k^* = \lvert\mathbf{p}^*_1-\mathbf{p}^*_2\rvert/2$ is the relative momentum defined in terms of the particle momenta $\mathbf{p}^*_1$ and $\mathbf{p}^*_2$ in the pair's rest frame. Here, $N_{\mathrm{same}}$ is the relative momentum distribution of particle pairs of interest measured within the same collision and therefore contains the correlation signal, while $N_{\mathrm{mixed}}$ is a reference distribution constructed by combining particles from different collisions following the standard event-mixing technique~\cite{ALICE:2022boj}, where only events of similar multiplicity and collision position in the beam direction are mixed. The normalization constant, $\mathcal{N}$, is chosen such that $C(k^*)$ is equal to unity in a kinematic region where the final-state interactions are expected to be negligible. Theoretically, the correlation function is described by the Koonin--Pratt formula $C(k^*) = \int \mathrm{d}^3r^* S(r^*)|\Psi(k^*,r^*)|^2$, where the source function, $S(r^*)$, denotes the distribution of the relative distance, $r^*$, between the two hadrons at emission, and $\Psi(k^*,r^*)$ is their relative wave function. The source distribution can be experimentally constrained by studying the correlation of pairs with known final-state interactions. In pp collisions at the LHC energies, it can be described by a Gaussian function common to all primordial hadron pairs, with an additional correction accounting for the exponential decay of short-lived resonances whose mean decay length, $c\tau$, is comparable to the source size~\cite{ALICE:2020ibs,ALICE:2023sjd,ALICE:2025aur,Mihaylov:2023pyl}. In the case of p--p pairs, the effective source distribution can also be parametrized with a Gaussian function, $S(r^*)=1/(4\pi r_\mathrm{0}^{2})^{3/2} \exp[-r^{*2}/(4 r_\mathrm{0}^{2})]$, where $r_{\mathrm{0}}$ denotes an effective source size in the sense that it includes the effects of short-lived resonances. 

The extension of the femtoscopy method to three-particle systems requires measuring the distributions $N_{\mathrm{same}}$ and $N_{\mathrm{mixed}}$ in terms of the hypermomentum $Q_3$, which is a Lorentz-invariant quantity defined from a pair relative four-momentum $q_{ij}^\mu = 2\ (m_j p_i^\mu - m_i p_j^\mu)/(m_i+m_j)$ as $Q_3 = \sqrt{-q_{12}^{2} - q_{23}^2 - q_{31}^2}$. The extension of the Koonin--Pratt formula to three-particle correlations is done in hyperspherical coordinates and expressed as
\begin{align}
C(Q_3) = \int \mathrm{d}\rho\ \mathrm{d}\Omega\  \rho^5\ S(\rho, \Omega)\ \big|\Psi(Q_3,\rho, \Omega)\big|^2 \ .
\label{eq:3B}
\end{align}
The source function, $S(\rho, \Omega)$, and the three-body scattering wave function, $\Psi(Q_3,\rho, \Omega)$, are expressed in terms of the hyperradius $\rho = \sqrt{\mathbf{x}^2 + \mathbf{y}^2}$, where $\mathbf{x}$ and $\mathbf{y}$ are the Jacobi coordinates of the system, and $\Omega$ encodes the remaining angular dependence~\cite{Nielsen:2001hbm,Marcucci:2019hml}. In this work, the Jacobi coordinates are defined as $\mathbf{x} = \mathbf{r}_2 - \mathbf{r}_1$ and $\mathbf{y} = \sqrt{4/3} \ [\ \mathbf{r}_3 - (\mathbf{r}_1 + \mathbf{r}_2)/2\ ]$, where $\mathbf{r}_1$, $\mathbf{r}_2$ and $\mathbf{r}_3$ are the position vectors of the three particles in the system. Under the assumption of independent emission, the source function for three protons can be parametrized by a Gaussian with an effective source radius $\rho_0$ as $S(\rho,\Omega) = 1/(\pi^3 \rho_0^6)\ \exp(-\rho^2/\rho_0^2)$. The following assumption links the three- and two-particle source descriptions: the underlying emission geometry that governs two-proton correlations also determines the typical interparticle separations within proton triplets. In this approach, the three-proton source size is related to the two-proton source size as $\rho_0 = 2r_{\mathrm{0}}$~\cite{Kievsky:2023maf}, and thus can be constrained by measuring the p--p correlation function.

In this work, the two-body p--p and three-body p--p--p correlation functions are measured in pp collisions at the center-of-mass energy of $\sqrt{s}=13.6$ TeV collected by the ALICE Collaboration at the LHC. The measurements were enabled by the upgraded ALICE apparatus, which operates in continuous data-taking mode as described in Appendix~\ref{sec:ExperimentalAparatus}~\cite{ALICE:2023udb,Buncic:2015ari}. The work analyzes the data recorded in the year 2024 during the Run 3 campaign. Only a small sample of minimum-bias (MB) data is recorded for analysis. While such a sample is sufficient to measure the two-proton correlation function, the three-proton measurement requires taking advantage of the full integrated luminosity acquired by the ALICE Collaboration, which corresponds to 53.1 pb$^{-1}$. For this purpose, three-body software triggers were developed which ensure that, in addition to an MB sample, all collisions containing triplets with low hypermomentum $Q_3<1.4\ \mathrm{GeV}/c$ are also recorded. The hypermomentum limit is justified as the correlation signal is expected in the low-energy region of $Q_3\lesssim1\mathrm{GeV}/c$ and necessary due to storage limitations. Such triplets are extremely rare and were found only once in every 30000 MB collisions. Since the trigger selection does not include the full phase-space of collisions with triplets due to the hypermomentum requirement, the $N_{\mathrm{mixed}}$ distribution is obtained from corresponding MB data of collisions that contain a triplet of interest with any $Q_3$ value. To improve statistical precision, particle and antiparticle correlation functions are combined, as the interaction is expected to be the same, and no differences in the measured correlation functions are observed. Protons and antiprotons are required to have transverse momentum in the range 0.5--2.0 GeV/$c$. They are identified based on the specific energy loss in the Time Projection Chamber (TPC) detector and the time-of-flight measurement enabled by the corresponding TOF detector. The purity for selected protons and antiprotons is estimated as described in Appendix~\ref{sec:EventAndTrackSelections} and is on average 97\%. The resulting particle sample includes both primary protons produced in the
collision and secondary protons from the weak decays of $\Lambda$ and $\Sigma^{+}$ hyperons, as well as misidentified particles and protons that are assigned to the wrong collision, in which they were not produced, by the reconstruction algorithm. These fractions are estimated in a data-driven way. The details on the ALICE upgrades, the software triggers, particle identification and selection, as well as purity and fraction extractions, are further described in Appendices~\ref{sec:ExperimentalAparatus}~and~\ref{sec:EventAndTrackSelections}.

\begin{figure}[!ht]
    \centering
    \includegraphics[width=0.65\textwidth]{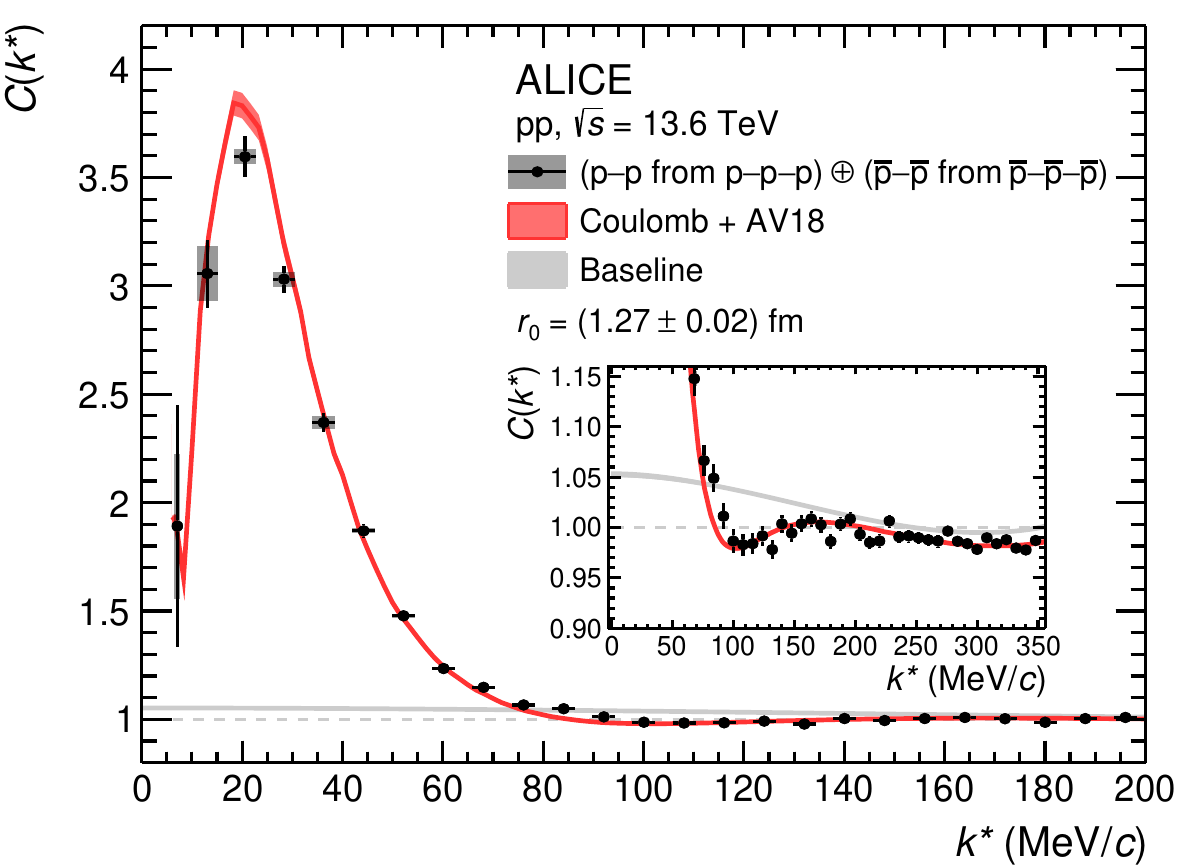}
    \caption{The p--p correlation function for proton pairs within p--p--p triplets measured in pp collisions at $\sqrt{s} = 13.6$ TeV is presented in black markers. The vertical lines represent the statistical uncertainty, while the shaded boxes correspond to the systematic uncertainty. The correlation function is normalized to unity in the range $k^* \in [150,200]$ MeV/$c$. The red band corresponds to the correlation function fit as described in the text. The gray band shows the baseline obtained from the fit, while the dashed gray line indicates unity.}
    \label{fig:ppCF}
\end{figure}
The first step of the analysis consists of studying the p--p correlation function, from which the effective source size for the triplets of interest and the non-femtoscopic correlations contributing to the three-proton system are determined.
The measured p--p correlation function obtained by combining all possible proton pairs within triplets is shown by the black markers in Fig.~\ref{fig:ppCF}. The vertical bars indicate the statistical uncertainties, while the shaded boxes represent the systematic uncertainties, which are treated as fully correlated from bin to bin (point-to-point) for all measurements presented in the following (see Appendix~\ref{sec:EventAndTrackSelections} for details on the data analysis). The horizontal black lines correspond to the statistical uncertainty in the determination of the mean $k^*$ for each bin. The measured signal includes not only correctly identified primary proton correlations (genuine) but also contributions from protons stemming from the weak decay of hyperons (feed-down), misidentifications, and protons from wrongly assigned collisions (see Appendix~\ref{sec:fits} for the modeling of the correlation function). The significant correlation signal at $k^* < 100$ MeV/$c$ has been extensively discussed in previous studies and originates from the two-proton final-state interaction~\cite{ALICE:2019buq,ALICE:2020ibs,Kievsky:2023maf,Gobel:2025afq,ALICE:2025wuy}. In this work, it is modeled by using the Argonne $v_{18}$ (AV18) nuclear potential~\cite{Wiringa:1994wb} and by additionally including the effects of Coulomb repulsion as well as the proper antisymmetrization of the two-proton wave function. Considering other interaction models leads to equally good descriptions of the correlation function at the per mille level since the resulting wave functions differ only at very small distances ($\leq$ 1 fm) as shown in Ref.~\cite{ALICE:2026tqh}. The main feed-down contribution comes from protons produced in weak decays of $\Lambda$ hyperons, which inherit information about the interaction between the parent particle and the other particle in the pair. Such residual p--$\Lambda$ contribution in this work is modeled using chiral EFT calculations~\cite{Haidenbauer:2019boi}. All modeled correlation signals include the effect of the experimental momentum resolution. Additionally, the measured correlation function may show effects from a non-femtoscopic background, which is usually attributed to long-range momentum correlations induced by energy-momentum conservation effects and can be well described by a polynomial baseline~\cite{Grosse-Oetringhaus:2024bwr,ALICE:2018ysd}. The resulting correlation function is then fitted to the data to extract the baseline parameters and the source size, which are required inputs for the interpretation of the measured three-body correlation function. The fit is performed in the region $k^* < 375$ MeV/$c$ and accounts for both the statistical and correlated systematic uncertainties, yielding a very good description of the data, with a $\chi^2/\mathrm{ndf} = 59.1/47 = 1.26$ (p-value = 0.11). The results are shown in Fig.~\ref{fig:ppCF} by a red band. The width of the band corresponds to the systematic uncertainty of the fit, and the full fitting procedure is described in Appendix~\ref{sec:fits}. The resulting baseline is represented by a light gray band, and the source size extracted from the fit of the data is $r_0 = (1.27 \pm 0.02)$ fm. The obtained source radius lies within the range of values expected from a recent differential study of the p--p source in MB collisions, which characterized its dependence on charged-track multiplicity and transverse mass~\cite{ALICE:2026tqh}. The radius uncertainty is evaluated accounting for both statistical and correlated systematic uncertainties in the data and the model. Following the relation between two- and three-proton source radii, the p--p--p source size becomes $\rho_0 = (2.54 \pm 0.04)$ fm.

\begin{figure}[!ht]
    \centering\includegraphics[width=0.65\textwidth]{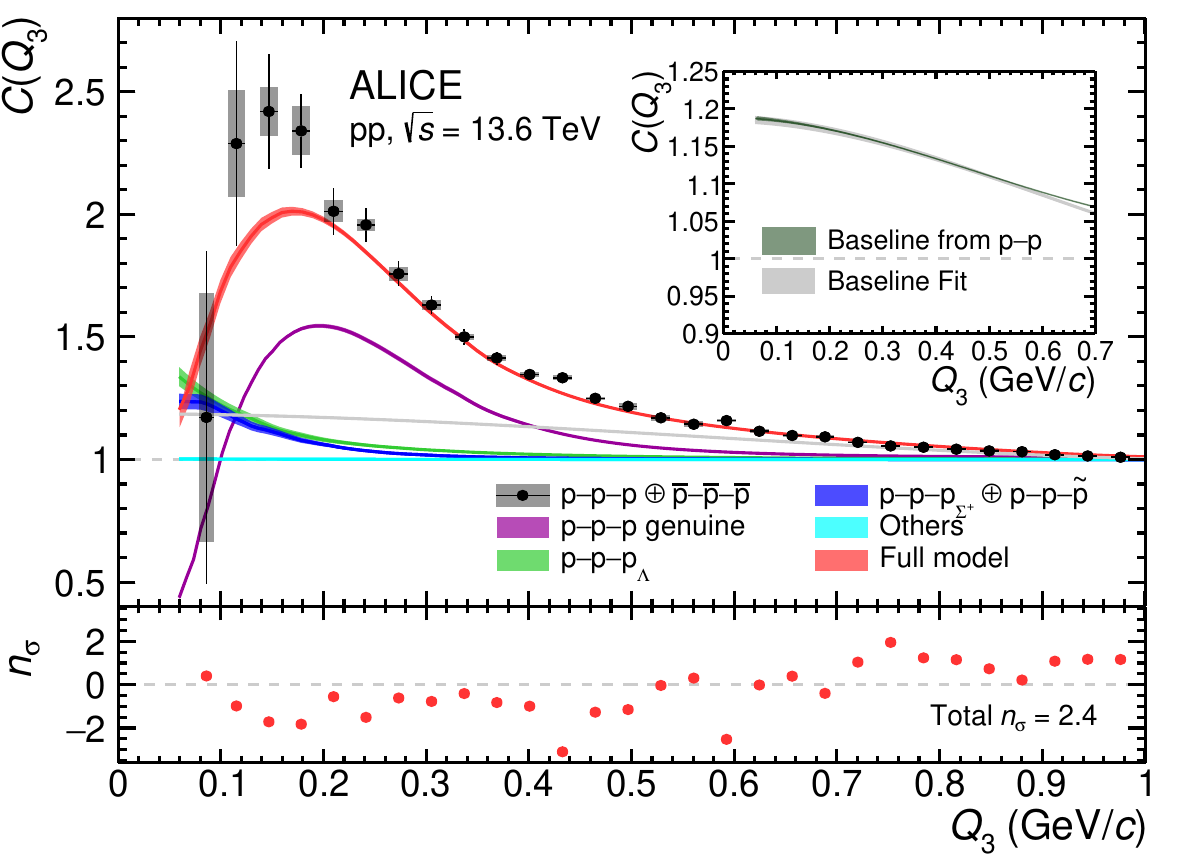}
    \caption{ The p--p--p correlation function measured in pp collisions at $\sqrt{s} = 13.6$ TeV, normalized to unity in the range $Q_3 \in [1.0,1.2]$ GeV/$c$. The vertical bars indicate the statistical uncertainties, while the shaded boxes represent the systematic ones. The magenta band represents the genuine p--p--p contribution, and the other colored bands correspond to the different components used in the modeling of the correlation function (see text for details). The red band is the result of the fit. The non-femtoscopic baseline is shown in light gray. The bottom panel reports the number of standard deviations ($n_\sigma$) between the data and the full model in each bin, obtained by combining the corresponding statistical and systematic uncertainties. The inset shows the comparison between the non-femtoscopic baseline obtained from the fit of the p--p--p correlation function (light gray band) and the corresponding result inferred from the two-body p--p measurement (darker green band).}
    \label{fig:ppANDpppCFs}
\end{figure} 

The measured p--p--p correlation function is shown in Fig.~\ref{fig:ppANDpppCFs} with black markers. It deviates from unity over a broad range of $Q_3$ below 1 GeV/$c$, indicating sensitivity to the hadronic final-state interaction. As in the two-body case, the measured signal includes not only the genuine correlation but also the feed-down contributions, misidentifications, and protons from wrongly assigned collisions. The full decomposition of the correlation function is given by
\begin{align}
   C(Q_3) = B(Q_3)\ &\big[\ \lambda_{\mathrm{ppp}}(Q_3)\ C_{\mathrm{ppp}}(Q_3)
+ \lambda_{\mathrm{ppp_\Lambda}}(Q_3)\ C_{\mathrm{ppp_\Lambda}}(Q_3)
+ \lambda_{\mathrm{ppp_{\Sigma^+}}}(Q_3)\ C_{\mathrm{ppp_{\Sigma^+}}}(Q_3)\ \nonumber \\ & + \lambda_{\mathrm{pp\tilde{p}}}(Q_3)\ C_{\mathrm{pp\tilde{p}}}(Q_3) + \lambda_\mathrm{others} (Q_3)\ C_{\mathrm{others}}(Q_3) \big] \ ,
\label{eq:decoPPP}
\end{align}
where $B(Q_3)$ denotes the non-femtoscopic baseline which has to be fitted to the data, while the terms labeled by the subscripts represent the different triplet contributions. The baseline $B(Q_3)$ accounts for possible distortions of the correlation function due to underlying events, such as mini-jet phenomena arising from hard processes at the parton level and long-range momentum correlations. It enters as a multiplicative factor, since the final-state interactions modify the correlation function on top of this underlying contribution. The relative weights, $\lambda_{ijk}(Q_3)$, are determined in a data-driven manner as the product of purities and fractions of each particle $i,j$ and $k$ in the triplet, as described in Appendix~\ref{sec:EventAndTrackSelections}. Here, p indicates correctly identified primary protons, p$_\mathrm{X}$ denotes a proton coming from a weak decay of particle X, and $\mathrm{\tilde{p}}$ are the misidentified or wrongly assigned protons. In the region $Q_3 < 1$ GeV/$c$, the dominant femtoscopic signal arises from genuine p--p--p correlations $C_{\mathrm{ppp}}(Q_3)$ (79\% of the triplets), followed by residual feed-down $C_{\mathrm{ppp_\Lambda}}(Q_3)$ (7.2\%) and $C_{\mathrm{ppp_{\Sigma^+}}}(Q_3)$ (3.5\%) contributions, as well as correlations involving a misidentified or incorrectly associated proton $C_{\mathrm{pp\tilde{p}}}(Q_3)$ (7.6\%). The remaining 2.7\% of the triplets contain only one or no primary protons. Such contributions are included in $C_{\mathrm{others}}(Q_3)$ and lead to negligible deviations from unity~($\leq 1.0016$). 

The $C_{\mathrm{ppp}}(Q_3)$ term is computed from the full three-body p--p--p scattering wave function calculations following Ref.~\cite{Garrido:2025lar}, using the source radius, $\rho_0$, fixed by the two-proton measurement described above. The two-nucleon interaction is modeled with the AV18 potential, whereas the Coulomb interaction is treated as in Refs.~\cite{Kievsky:2024cwb,Kievsky:2023maf}. The three-nucleon force is not considered, since in the p--p--p system it is suppressed by Pauli blocking and its effect on the correlation function is smaller than 0.1\%~\cite{Kievsky:2023maf,Garrido:2025lar}. A different choice for the nuclear potential is discussed in Appendix~\ref{sec:Norfolk}, and negligible differences, below 1.5\%, in the peak region are found in the modeled correlation function. The resulting $C_{\mathrm{ppp}}(Q_3)$ component is represented by the dark magenta band in Fig.~\ref{fig:ppANDpppCFs}, already scaled by the corresponding $\lambda_\mathrm{ppp}(Q_3)$ parameter and corrected to account for the experimental momentum resolution. The widths of all bands in Fig.~\ref{fig:ppANDpppCFs} correspond to the uncertainties in the modeling, evaluated from the fit procedure and including systematic variations of every contribution as described below. In the case of $C_{\mathrm{ppp}}(Q_3)$, it is dominated by the variation of the source radius $\rho_0$ within the measured uncertainty. Additionally, the variation of $\lambda$ parameters for feed-down and misidentified or incorrectly associated protons by $\pm10\%$ relative to their nominal values is included to account for uncertainties in the determination of the secondary fractions from Monte Carlo template fits, while $\lambda_{\mathrm{ppp}}(Q_3)$ was determined by requiring that the sum of all $\lambda$ parameters remains equal to unity.
The correlation functions associated with feed-down, misidentified contributions, and non-femtoscopic effects are constrained by data-driven procedures. The residual feed-down $C_{\mathrm{ppp_\Lambda}}(Q_3)$, shown in Fig.~\ref{fig:ppANDpppCFs} with a green band, is experimentally constrained from the p--p--$\Lambda$ correlation function, evaluated using the $\Lambda$ selection criteria described in Appendix~\ref{sec:EventAndTrackSelections}, and corrected for the fraction of correctly identified triplets with primary protons and for the momentum smearing induced by the $\Lambda \rightarrow~ $p$ \pi$ decay kinematics. The $1\sigma$ experimental uncertainty, and a $\pm10\%$ variation of the relevant triplet fraction are included in the fitting procedure.
All remaining components in Eq.~\ref{eq:decoPPP} are evaluated with the cumulant expansion method, which constructs a three-particle correlation function from two-particle inputs~\cite{DelGrande:2021mju}. The relevant pairwise contributions can be inferred from the p--p analysis and include the genuine p--p as well as the p--$\Lambda$ feed-down, while the remaining correlation sources can be neglected. Accordingly, the following contributions to the three-body correlation function must be considered: (p--p--p$_{\Sigma^+} \oplus$ p--p--$\tilde{\mathrm{p}}$), (p--p$_\Lambda$--p$_\Lambda$), and (p--p$_\Lambda$--p$_{\Sigma^+} \oplus$ p--p$_\Lambda$--$\tilde{\mathrm{p}}$), where the grouped triplets are modeled using the same pairwise interactions. Since only (p--p--p$_{\Sigma^+} \oplus$ p--p--$\tilde{\mathrm{p}}$) triplets lead to a sizable deviation from unity (blue band), driven by the interaction of the p--p pairs, the remaining two contributions are absorbed into the flat component $C_\mathrm{others}(Q_3)$ (cyan band). The uncertainty of (p--p--p$_{\Sigma^+} \oplus$ p--p--$\tilde{\mathrm{p}}$) contribution includes a variation of the corresponding correlation function strength by $\pm 20\%$ to account for the assumptions on the interactions made in the cumulant method. Finally, the non-femtoscopic baseline is parametrized consistently with the two-body case by a third-order polynomial without a linear term,
$B(Q_{3}) = a + b(Q_3)^2 + c(Q_3)^3$, where the coefficients $a$, $b$, and $c$ are determined from a fit to the measured p--p--p correlation function in the range $Q_3 \in [0,1.6]$ GeV/$c$, including both the statistical and correlated systematic uncertainties. Since the offline triggers retain only triplets with $Q_3 <$ 1.4 GeV/$c$, the triggered sample is distorted near this limit and is therefore unsuitable for constraining the baseline at large hypermomenta. For this reason, in the region $Q_3 \in [1.1,1.6]$ GeV/$c$ the correlation function is instead evaluated on the MB sample, which is free of this selection. The resulting baseline is shown as a light gray band in Fig.~\ref{fig:ppANDpppCFs}. The uncertainty includes variation on the $Q_3$ range, at which the correlation function is obtained using MB data by $\pm 100$ MeV/$c$ at the lower limit, and also the resulting fit uncertainties on the $a,b,$ and $c$ parameters.
Since the non-femtoscopic effects are dominated by pairwise contributions, the three-body baseline can also be estimated via the cumulant expansion method using the extracted two-body baseline as input~\cite{DelGrande:2021mju}. A good agreement is found between the two methods, as shown in the inset in Fig.~\ref{fig:ppANDpppCFs}. This comparison validates the description of the three-body correlation function: the baseline extracted from the three-body fit is consistent with the expectation inferred from the two-body measurement, confirming that it reflects genuine two-body effects rather than compensating for limitations in the three-body theory. Since the baseline is the only quantity fitted in the three-body case, and its behavior is independently confirmed in a data-driven way by the two-body measurement, the remaining agreement between model and data constitutes a direct comparison to the interaction models.

The full modeled correlation function corresponding to Eq.~\ref{eq:decoPPP} is depicted as a red band. Agreement with the data is quantified in the bottom panel in terms of standard deviations, $n_\sigma$, obtained by combining statistical and systematic uncertainties from both the data and the model. The overall deviation amounts to $n_\sigma = 2.4$ for $Q_3 < 1$ GeV/$c$, and to $n_\sigma = 1.5$ for $Q_3 < 0.25$ GeV/$c$. Such $n_\sigma$ values are obtained from the p-value of the total $\chi^2$ in the considered $Q_3$ intervals, by including the statistical and correlated systematic uncertainties of the data and the model. The very good agreement with the full three-body calculations, which treat the process as a 3$\rightarrow$3 reaction, indicates that such a measurement provides access to the three-proton continuum. It is important to note that the Coulomb interaction in the p--p--p system is treated approximately, as no exact analytic solution exists for the three-body Coulomb problem. Future improvements in the theory, including also other approximations used~\cite{Garrido:2025lar}, could lead to even a better description of the data. Moreover, the feed-down correlations can be constrained even better in the future if corresponding calculations are developed or corresponding measurements become available. In the scope of the present work, the next step is to investigate the sensitivity of the correlation function to the different components of the interaction.

\section{Genuine three-baryon correlation function}

\begin{figure}[h]
    \centering
    \includegraphics[width=0.85\textwidth]{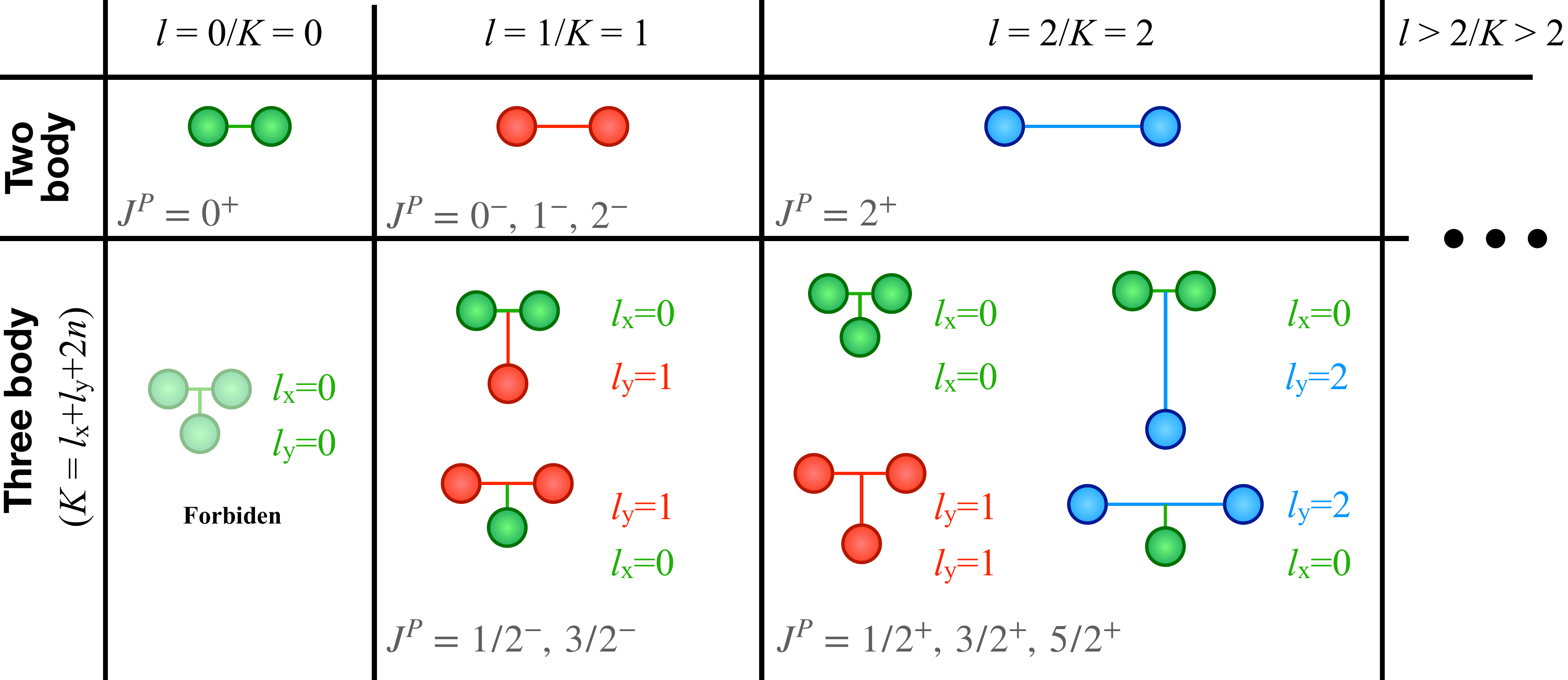}
    \caption{Possible configurations of protons in two- and three-body systems shown for correspondence to their quantum numbers, the angular momentum, $l$, for the two-body system and the grand-angular momentum, $K$, for the three-body system. The $K=0$ configuration is forbidden for three protons due to antisymmetrization arguments. The green color here encodes particles in $l=0$, red in $l=1$ and blue in $l=2$ states.}
    \label{fig:configurations}
\end{figure}

The genuine p--p--p correlation function $C_{\mathrm{ppp}}(Q_3)$ is isolated using Eq.~\ref{eq:decoPPP} by subtracting the experimentally constrained residual and non-femtoscopic contributions from the measured correlation function. The effects of momentum resolution are below 1\% for $Q_3 > 0.115$ GeV/$c$ and reach at most 3\% in the lowest measured kinematic range, which is negligible compared to the statistical and systematic uncertainties and thus no additional corrections are applied to the extracted genuine correlation function. The resulting observable can be compared directly to the theoretical expectation from Eq.~\ref{eq:3B} for a Gaussian source with $\rho_0 = (2.54 \pm 0.04)$~fm. The three-proton scattering problem in the continuum is solved within the hyperspherical adiabatic expansion~\cite{Nielsen:2001hbm}. The three-body wave function is decomposed as an infinite sum of terms of definite total angular momentum and parity, $J^P$, which depend on the angular momentum between two given particles, $l_x$, and the relative angular momentum of the third particle with the pair, $l_y$. Such expansion is usually ordered by the grand-angular momentum, $K$, which sets the allowed configurations of the angular momentum as $K=l_x + l_y +2 n$, where $n$ is an integer quantum number related to the order of the hyperspherical polynomial. The ordering of the expansion in $K$ for the construction of the three-body wave function is shown in the second row in Fig.~\ref{fig:configurations} with all the possible configurations in $l_x$, $l_y$ and $n$ for a given $K$.
A similar ordering is used in the two-body case, where the orbital angular momentum, $l$, between the pair determines the expansion in partial waves as shown in the first row in Fig.~\ref{fig:configurations}. The possible $J^P$ states for p--p and p--p--p systems are also listed at each order of the expansion, and they are the quantum numbers for which the interaction is evaluated. The wave function of the p--p system is usually constructed by including the nuclear interaction in the lower orders of the expansion in $l$, corresponding to $s$-wave ($l=0$), $p$-wave ($l=1$) and $d$-wave ($l=2$). For larger angular momenta, the effects of the strong interaction become subdominant because of the increasing effect of the centrifugal barrier, an effective repulsive contribution that arises from the angular part of the kinetic-energy term in the Hamiltonian and scales as $l(l+1)$. Thus, such higher orders in the expansion include only the Coulomb and the centrifugal barrier effects. In the three-body case, the expansion in $K$ does not uniquely identify the relative angular momenta of the pairs in the triplets as shown in Fig.~\ref{fig:configurations}. In particular, one can see that the $s$-wave NN interaction in $l_x=0$ appears in all the orders in $K$ (green lines in the sketch). The $p$-wave interaction, instead, contributes starting from $K=1$ (red lines) and the $d$-wave interaction from $K=2$ (blue lines). For the measured $Q_3$ interval, the full p--p--p wave function is obtained by considering the nuclear interaction in all two-body partial waves ($s$, $p$, $d$) and in the expansion up to $K=7$, as the effect of the NN interaction becomes negligible for higher values of the grand-angular momentum~\cite{Garrido:2025lar}. In this work, the sensitivity of the p--p--p correlation function to the NN interaction is investigated by comparing the data to the correlation functions calculated under two controlled assumptions: (i) switching off the NN interaction in a selected pair partial wave (relative angular-momentum channel) within the triplet, and (ii) progressively truncating the nuclear-interaction contribution at different orders in the expansion in $K$.

\begin{figure}[h]
    \centering
    \includegraphics[width=0.48\textwidth]{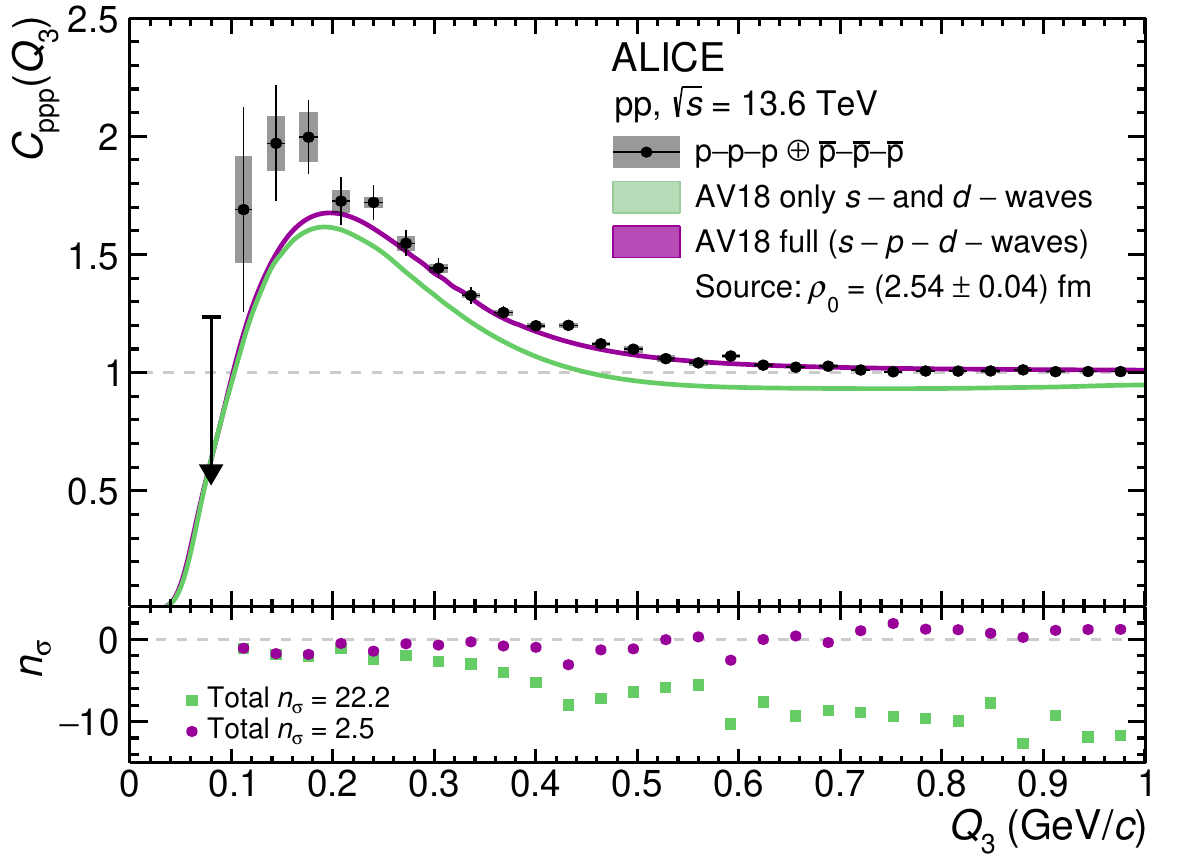}
    \includegraphics[width=0.48\textwidth]{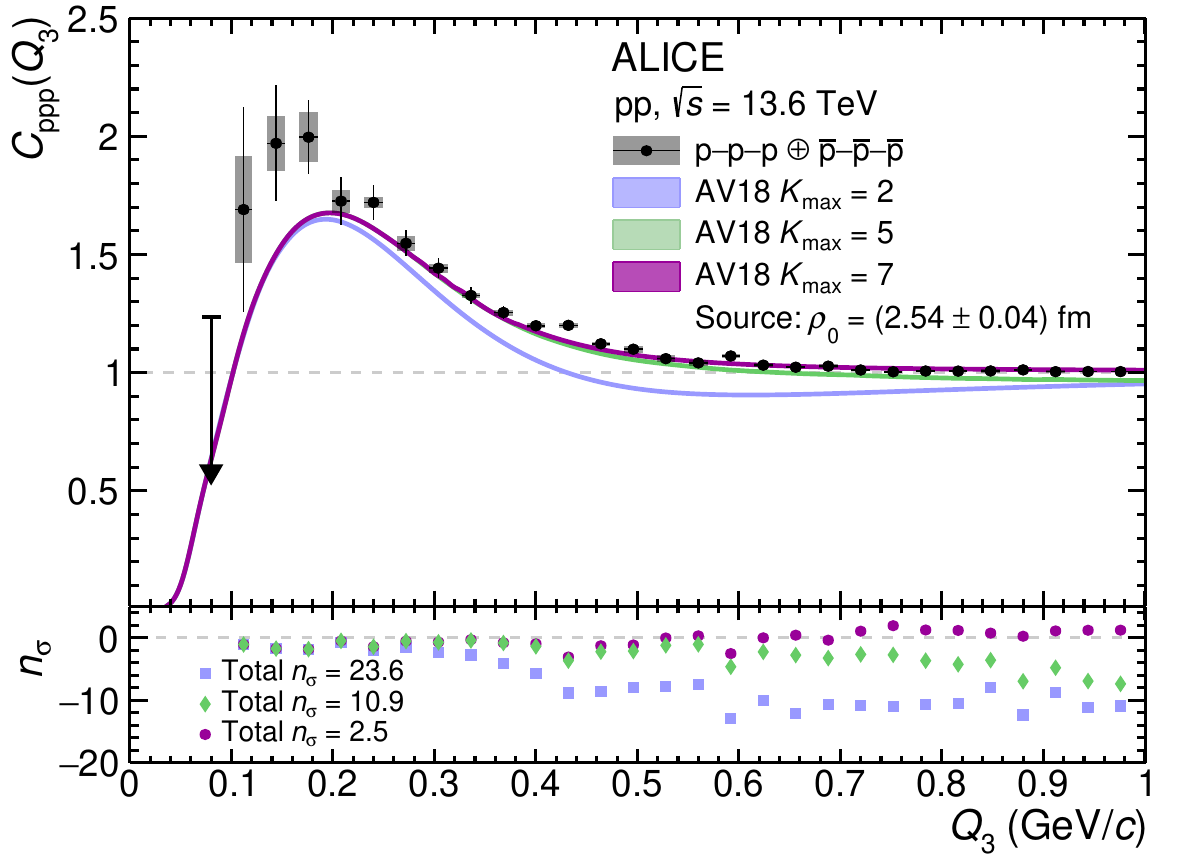}
    \caption{Genuine p--p--p correlation function with statistical (bars) and systematic uncertainties
(boxes). Left: The magenta and green bands depict the calculations from Ref.~\cite{Garrido:2025lar} obtained by using the AV18 nuclear potential with and without including the $p$-wave component of the interaction, respectively. Calculations are performed using $K_\mathrm{max}=7$. Right: The different color bands correspond to calculations including the interaction in \textit{s}-, \textit{p}-, and \textit{d}-waves of AV18 potential, and differ by the maximum value of the grand-angular momentum $K$ used to construct the interacting part of the scattering wave function: $K_\mathrm{max}=2$ (blue band), $K_\mathrm{max}=5$ (green band) and $K_\mathrm{max}=7$ (magenta band). The magenta band is identical in both panels. For all bands, the width corresponds to the uncertainty on the source radius. The bottom panel reports the number of standard deviations ($n_\sigma$) between the theory and the data in each bin, evaluated by combining the statistical and systematic uncertainties of the data and the uncertainty of the source radius used in the calculations. The total $n_\sigma$ values refer to the full $Q_3$ range shown in the plots. The first bin is reported as a 1$\sigma$ upper limit, as the corresponding confidence interval extends into the non-physical region (values below zero).}
    \label{fig:CFs1}
\end{figure}

The measurement is compared to calculations in the left panel of Fig.~\ref{fig:CFs1}, including all two-body partial waves (magenta band) and to a variant in which the nuclear interaction in the $p$-wave is removed (green band). Following the sketch in Fig.~\ref{fig:configurations}, the removal of the $p$-wave interaction is implemented by treating the $p$-wave channels (red diagrams) as a free wave at all orders in the expansion in $K$. The corresponding bin-by-bin $n_\sigma$ values are shown in the lower panel. Excluding the nuclear interaction in the $p$-wave leads to a pronounced disagreement with the data, with a total statistical significance of $22.2\sigma$, whereas the full calculation reproduces the measurement with an overall agreement of $2.5\sigma$. These results demonstrate that the p--p--p correlation function is sensitive to the same partial-wave dynamics of the NN interaction that are constrained by scattering data. The agreement of the three-body calculations with the p--p--p measurement additionally provides the first direct experimental benchmark for a three-nucleon wave function in the isospin 3/2 configuration. Previously, only isospin 1/2 systems could be studied directly through nucleon--deuteron scattering.

The expansion in orders of $K$ used to construct the three-body wave function can be truncated at a value $K_\mathrm{max}$ above which all waves are treated as free of the nuclear force. Referring to the sketches in Fig.~\ref{fig:configurations}, choosing $K_\mathrm{max}=1$ would mean that the nuclear $s$- and $p$-wave interactions are included for $K=1$, whereas for $K>1$, the nuclear force is disabled in all configurations. Similarly to the two-body case, the value of $K_\mathrm{max}$ must be chosen such that the nuclear interaction at $K > K_\mathrm{max}$ is subdominant with respect to the centrifugal barrier. As the centrifugal barrier grows with the grand-angular momentum according to $(K+3/2)(K+5/2)$, the contribution of the nuclear interaction in the three-proton system would be expected to be quickly suppressed with increasing $K$, similarly to the behavior observed in two-body systems. 
In the latter case, the convergence to the free system is intuitive because the NN interaction becomes progressively less important with increasing angular momentum of the pair, eventually leading to a regime where the system can be approximated as effectively non-interacting. To study this effect in the three-body case, the measured p--p--p correlation function is compared to the theory predictions with different $K_\mathrm{max}$ assumptions as shown in the right panel of Fig.~\ref{fig:CFs1}. The calculation corresponding to $K_\mathrm{max} = 2$ (blue band) describes the peak observed in the low $Q_3$ range. The agreement with the data is quantified in units of standard deviations, resulting in $n_\sigma < 2$ up to 0.2 GeV/$c$. However, it fails to describe the data at larger $Q_3$ values. The inclusion of the interaction up to $K\leq 5$ (green band) introduces additional attraction, yet the description remains unsatisfactory. Interestingly, the number of $J^P$ states included for $K_\mathrm{max} = 5$ is the same as that needed to successfully describe the p--d correlation function~\cite{ALICE:2023bny}. Finally, including the nuclear interaction up to $K\leq 7$ results in a good agreement with the data, quantified as $n_\sigma = 2.5$. In contrast to the two-body case, the contribution from the nuclear interaction remains significant up to a high value of $K_\mathrm{max}$. Theoretical studies confirm that, in the measured $Q_3$ range, the nuclear force remains relevant up to $K_\mathrm{max}=7$~\cite{Garrido:2025lar}. In the three-body dynamics, increasing values of the grand-angular momentum accommodate multiple configurations of the particle pairs, including those in which one pair remains in a relative $l_x=0$ state. As a consequence, the nuclear interaction contributes significantly even at large $K$, making the convergence of the calculations inherently slower than in the two-body case. The right panel of Fig.~\ref{fig:CFs1} also indicates that larger values of $K_\mathrm{max}$ are needed to reproduce the correlation function over an increasingly larger $Q_3$ range.

At large $Q_3$, the kinetic energy of the particles becomes dominant over the interaction potentials, and the system is expected to approach the asymptotic regime where particles are sufficiently separated to be treated as freely moving. With the increasing energy, only long-range effects such as Coulomb repulsion and quantum statistics should play a role and would drive the correlation function below unity before reaching one. However, the measured correlation function remains above unity as it approaches one, as seen in Fig.~\ref{fig:CFs1}, indicating the presence of an effective long-range attraction. Although the only attractive interaction in the p--p--p system is the short-range two-body nuclear force, this effect emerges naturally from the three-body dynamics: at large $Q_3$, the system never reaches asymptotic configurations because two of the three protons can always remain close enough to interact. This results in an effective long-range attraction scaling as $\zeta/\rho^3$ at large values of $\rho$ (see Refs.~\cite{Higgins:2020pbe, Higgins:2020avy}). The constant $\zeta$ is proportional to the two-body scattering length, linking the present measurement to the universal physics of systems characterized by large scattering lengths~\cite{Braaten:2004rn,Kievsky:2021ghz}. Further details are given in Appendix~\ref{sec:convergence}. While previously discussed primarily from a theoretical perspective, the p--p--p correlation function measurement presents the first experimental signature of such effective long-range attraction in a three-body system. The observation of these distinctive three-body effects in the measured correlation function establishes femtoscopy as a powerful probe of three-body dynamics.

\section{Conclusions}
In summary, this work presents the measurement of the three-proton correlation function in pp collisions at $\sqrt{s}$ = 13.6 TeV, establishing three-baryon femtoscopy as an effective 3$\rightarrow$3 scattering experiment and providing the first direct access to the isospin $3/2$ three-body channel. A new, data-driven analysis strategy isolates the genuine p--p--p correlation function from background channels and experimental effects, enabling a direct comparison with state-of-the-art three-body continuum calculations. Such a comparison employing realistic nucleon--nucleon potentials shows that the measurement is sensitive to the partial-wave structure of the interaction: removing the $p$-wave contribution leads to a 22.2$\sigma$ disagreement with the data, while the complete calculation reproduces the measurement within 2.5$\sigma$. Two distinctive signatures of the three-unbound-nucleon system further emerge from the data. First, the nuclear interaction remains relevant up to high grand-angular momenta ($K\leq7$), in contrast to the rapid convergence of the two-body partial-wave expansion. Second, the correlation function stays above unity as it approaches the uncorrelated limit, providing the first experimental signature of an effective long-range attraction in a three-body continuum system. Both effects share a common origin in configurations where one proton is far from the other two, which remain close enough to interact, so that the short-range nuclear force continues to act even at large hyperradius. This long-range effect reflects the universal physics of systems with large two-body scattering lengths~\cite{Braaten:2004rn,Kievsky:2021ghz}. These results contribute to ongoing efforts to understand the possible quasi-bound states in three- and four-neutron systems~\cite{Kisamori:2016jie,Huang:2026sgc,PhysRevLett.133.012501}. Indeed, the numerical techniques used here are the same as those applied in Refs.~\cite{Higgins:2020pbe,Higgins:2020avy} to study the three- and four-neutron continuum in their lowest partial waves, which found an enhancement of the density of states at low energies that can account for the observation of few-neutron structures in that region. Furthermore, at the LHC energies, the copious production of protons and strange baryons makes it possible to apply the same method to systems such as p--p--$\Lambda$, p--p--$\Sigma$, and p--p--$\Xi$, opening a new program of precision studies of three-hadron interactions in the continuum that are otherwise experimentally out of reach. Such measurements are necessary for a deeper understanding of the hypernuclear structure and for constraining the equation of state of dense matter.

\newenvironment{acknowledgement}{\relax}{\relax}
\begin{acknowledgement}
\section*{Acknowledgements}
%

The ALICE Collaboration would like to thank all its engineers and technicians for their invaluable contributions to the construction of the experiment and the CERN accelerator teams for the outstanding performance of the LHC complex.
The ALICE Collaboration gratefully acknowledges the resources and support provided by all Grid centres and the Worldwide LHC Computing Grid (WLCG) collaboration.
The ALICE Collaboration acknowledges the following funding agencies for their support in building and running the ALICE detector:
A. I. Alikhanyan National Science Laboratory (Yerevan Physics Institute) Foundation (ANSL), State Committee of Science and World Federation of Scientists (WFS), Armenia;
Austrian Academy of Sciences, Austrian Science Fund (FWF): [M 2467-N36] and Nationalstiftung f\"{u}r Forschung, Technologie und Entwicklung, Austria;
Ministry of Communications and High Technologies, National Nuclear Research Center, Azerbaijan;
Rede Nacional de Física de Altas Energias (Renafae), Financiadora de Estudos e Projetos (Finep), Funda\c{c}\~{a}o de Amparo \`{a} Pesquisa do Estado de S\~{a}o Paulo (FAPESP) and The Sao Paulo Research Foundation  (FAPESP), Brazil;
Bulgarian Ministry of Education and Science, within the National Roadmap for Research Infrastructures 2020-2027 (object CERN), Bulgaria;
Ministry of Education of China (MOEC) , Ministry of Science \& Technology of China (MSTC) and National Natural Science Foundation of China (NSFC), China;
Ministry of Science and Education and Croatian Science Foundation, Croatia;
Centro de Aplicaciones Tecnol\'{o}gicas y Desarrollo Nuclear (CEADEN), Cubaenerg\'{\i}a, Cuba;
Ministry of Education, Youth and Sports of the Czech Republic, Czech Republic;
The Danish Council for Independent Research | Natural Sciences, the VILLUM FONDEN and Danish National Research Foundation (DNRF), Denmark;
Helsinki Institute of Physics (HIP), Finland;
Commissariat \`{a} l'Energie Atomique (CEA) and Institut National de Physique Nucl\'{e}aire et de Physique des Particules (IN2P3) and Centre National de la Recherche Scientifique (CNRS), France;
Bundesministerium f\"{u}r Forschung, Technologie und Raumfahrt (BMFTR) and GSI Helmholtzzentrum f\"{u}r Schwerionenforschung GmbH, Germany;
National Research, Development and Innovation Office, Hungary;
Department of Atomic Energy Government of India (DAE), Department of Science and Technology, Government of India (DST), University Grants Commission, Government of India (UGC) and Council of Scientific and Industrial Research (CSIR), India;
National Research and Innovation Agency - BRIN, Indonesia;
Istituto Nazionale di Fisica Nucleare (INFN), Italy;
Japanese Ministry of Education, Culture, Sports, Science and Technology (MEXT) and Japan Society for the Promotion of Science (JSPS) KAKENHI, Japan;
Consejo Nacional de Ciencia (CONACYT) y Tecnolog\'{i}a, through Fondo de Cooperaci\'{o}n Internacional en Ciencia y Tecnolog\'{i}a (FONCICYT) and Direcci\'{o}n General de Asuntos del Personal Academico (DGAPA), Mexico;
Nederlandse Organisatie voor Wetenschappelijk Onderzoek (NWO), Netherlands;
The Research Council of Norway, Norway;
Pontificia Universidad Cat\'{o}lica del Per\'{u}, Peru;
Ministry of Science and Higher Education, National Science Centre and WUT ID-UB, Poland;
Korea Institute of Science and Technology Information and National Research Foundation of Korea (NRF), Republic of Korea;
Ministry of Education and Scientific Research, Institute of Atomic Physics, Ministry of Research and Innovation and Institute of Atomic Physics and Universitatea Nationala de Stiinta si Tehnologie Politehnica Bucuresti, Romania;
Ministerstvo skolstva, vyskumu, vyvoja a mladeze SR, Slovakia;
National Research Foundation of South Africa, South Africa;
Swedish Research Council (VR) and Knut \& Alice Wallenberg Foundation (KAW), Sweden;
European Organization for Nuclear Research, Switzerland;
Suranaree University of Technology (SUT), National Science and Technology Development Agency (NSTDA) and National Science, Research and Innovation Fund (NSRF via PMU-B B05F650021), Thailand;
Turkish Energy, Nuclear and Mineral Research Agency (TENMAK), Turkey;
National Academy of  Sciences of Ukraine, Ukraine;
Science and Technology Facilities Council (STFC), United Kingdom;
National Science Foundation of the United States of America (NSF) and United States Department of Energy, Office of Nuclear Physics (DOE NP), United States of America.
In addition, individual groups or members have received support from:
Czech Science Foundation (grant no. 23-07499S), Czech Republic;
FORTE project, reg.\ no.\ CZ.02.01.01/00/22\_008/0004632, Czech Republic, co-funded by the European Union, Czech Republic;
European Research Council (grant no. 101220549), European Union;
Deutsche Forschungs Gemeinschaft (DFG, German Research Foundation) ``Neutrinos and Dark Matter in Astro- and Particle Physics'' (grant no. SFB 1258), Germany;
CONVECS project, CUP C97H23001700002 FESR 2021-2027 program, Italy;
Ministerio de Ciencia, Innovaci\'on y Universidades, Agencia Estatal de Investigaci\'on, Spain (Grant No. PID2022-136992NB-I00).

\end{acknowledgement}

\bibliographystyle{utphys}   
\bibliography{bibliography}

\newpage
\appendix
\section{Appendix} 

 \subsection{Experimental apparatus and data samples}\label{sec:ExperimentalAparatus}

The analyzed pp collisions were recorded at the center-of-mass energy
$\sqrt{s}=13.6$~TeV by the ALICE Collaboration at the LHC during LHC Run 3 in 2024. The ALICE detector underwent a major upgrade during the second LHC long shutdown (2019–2021)~\cite{ALICE:2023udb}. The upgrades relevant to the present measurement include the installation of a new Inner Tracking System (ITS) as the innermost detector~\cite{ALICE:2013nwm}, the replacement of the Time Projection Chamber (TPC) multi-wire proportional chambers with gas electron multiplier readout chambers~\cite{ALICETPC:2020ann}, and the installation of new readout electronics for the Time-Of-Flight (TOF) detector~\cite{ALICETOF:2025tof}. The new ITS is made of seven layers of silicon sensors based on monolithic active pixel technology~\cite{ALICE:2013nwm}, enabling precise determination of track parameters close to the interaction point and, consequently, accurate reconstruction of primary and secondary vertices. The TPC provides up to 152 space points for charged-particle tracking and delivers particle identification via measurements of the specific ionization energy loss, d$E$/d$x$. At higher momenta, where the separation power of the TPC begins to decrease, particle identification is complemented by the TOF detector, which measures the time of flight of charged particles from the interaction point. In addition, the Fast Interaction Trigger (FIT) suite~\cite{Trzaska:2017reu}, comprising the FV0, FT0, and FDD sub-detectors, was installed at forward and backward rapidities and provides collision time, luminosity monitoring and centrality determination in heavy-ion collisions. With these upgrades deployed ahead of LHC Run 3, the ALICE detector can be operated in continuous data-taking mode~\cite{ALICE:2023udb,Buncic:2015ari}. The new readout system supports pp interaction rates of 660 kHz and above, far exceeding those achieved in Run 2, while preserving excellent tracking and particle-identification performance. The data are recorded in time frames of $\approx$3 milliseconds and include all detector responses.

The recorded data are kept for analysis following two strategies: a subsample of the recorded minimum-bias data, and the events chosen by the offline trigger selection (OTS). The OTS is developed to select events identified as relevant for specific analyses and is applied at the software level within the computing framework after full calibration and reconstruction based on physics criteria~\cite{Buncic:2015ari,ALICE-PUBLIC-2020-005}. The dedicated event-skimming strategy for the presented measurement, which also allows sufficient margin for systematic studies, first identifies the proton candidates, then forms p--p--p triplets produced in the same events and selects only events with at least one triplet with reconstructed hypermomentum $Q_3$ less than 1.4 GeV/$c$. The same strategy was applied in p--p--$\Lambda$ analysis, which, as mentioned in Section~\ref{sec:analysis}, is performed to estimate feed-down to the presented p--p--p correlation function measurement. The skimmed data are analyzed to obtain the same-event distribution. However, the hypermomentum requirement alters the selected event phase space, and therefore, the skimmed data cannot be used to construct the mixed-event distribution. Instead, the corresponding minimum-bias sample is used. The same selection criteria are then applied at the analysis level to both samples.

 \subsection{Collision and track selections}\label{sec:EventAndTrackSelections}
The analyzed collisions were required to have a valid FT0 time signal consistent with the LHC bunch-crossing time. Only collisions with a reconstructed primary vertex within 10 cm of the nominal position along the beam axis were selected. The proton and $\Lambda$ candidates were selected using topological, kinematic, and track-quality criteria, which were later varied to evaluate the associated systematic uncertainties. In the following, the corresponding systematic variations are given in parentheses. Primary proton candidates were required to satisfy $0.5 < p_{\mathrm{T}} < 2~\mathrm{GeV}/c$ and $|\eta| < 0.8\ (0.77, 0.83)$. Particle identification (PID) was performed by comparing the measured TPC and TOF signals to the expected response for protons. The level of agreement is expressed in units of the detector resolution, $n_{\sigma}^{\mathrm{PID}}$. For protons with $p_{\mathrm{T}} < 0.75~\mathrm{GeV}/c$, $n_{\sigma}^{\mathrm{PID}}$ was evaluated using the TPC only. At higher transverse momenta, the TPC and TOF information was combined as $n_{\sigma}^{\mathrm{PID}}=\sqrt{n_{\sigma,\mathrm{TPC}}^{2}+n_{\sigma,\mathrm{TOF}}^{2}}$. In both cases, a selection of $n_{\sigma}^{\mathrm{PID}} < 3$ (2.5, 3.5) was applied. To improve the track quality, only tracks with at least 90 (110) out of 152 associated space points (clusters) and at least 90 (80, 110) associated crossed rows in the TPC are accepted. TPC clusters can be shared across tracks, so a maximum shared-cluster requirement can be applied. Since no significant change in the correlation function shape was observed, it was not included in the main selection to maximize the statistical sample size, but it was considered as a systematic variation requiring a maximum of five shared clusters. A minimum requirement of 5 (0, 7) out of 7 hits in the ITS is also imposed, with at least 1 (0, 3) in any of the three innermost layers, to ensure good pointing resolution. To reject particles stemming from weak decays, the distance of closest approach (DCA) to the primary vertex (PV) is constrained in both the transverse plane and along the beam axis. Due to the high PV resolution that is achieved with the new ITS detector, a strict transverse-momentum-dependent selection is applied such that $\left| \mathrm{DCA}_{xy} \right| < \left(0.004 + 0.013\,\frac{\mathrm{GeV}/c}{p_{\mathrm T}}\right)\,\mathrm{cm}$, which was optimized based on the $\mathrm{DCA}_{xy}$ resolution. For the systematic variations, an additional (fixed) selection was applied, requiring $\mathrm{DCA}_{xy} < 0.1~\mathrm{cm}$.
The $|\mathrm{DCA}_{z}|$ was required to be less than $0.2~\mathrm{cm}$ $(0.05~\mathrm{cm})$.

\begin{figure}
    \centering
    \includegraphics[width=0.65\linewidth]{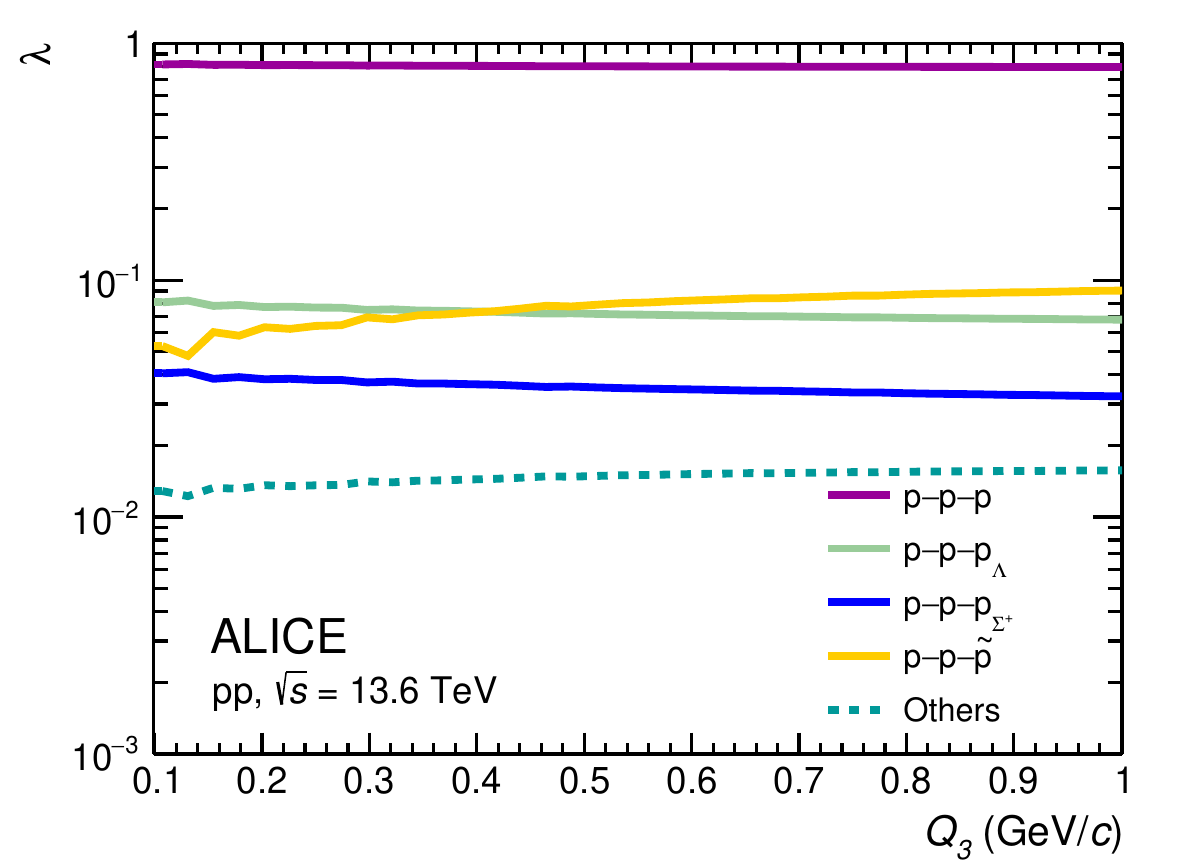}
    \caption{$\lambda(Q_3)$ parameters for the relevant components used in the decomposition of the p--p--p correlation function in Section~\ref{sec:analysis}.}
    \label{fig:lambdaparam}
\end{figure}

The purity of selected proton candidates is estimated with a data-driven method, by fitting the $n_{\sigma}^{\mathrm{PID}}$ distribution. The fits are performed in $p_{\mathrm{T}}$ bins, and the extracted purity as a function of $p_{\mathrm{T}}$ is later used to obtain $Q_3$ dependent $\lambda$ parameters. The purity decreases from 99.5\% to 92\% with increasing \pT. The selected, correctly identified protons are not only primary particles, but can also originate from weak decays of strange baryons, interactions in the detector material, or misassociation with the wrong collision. To estimate these fractions, the measured DCA distributions are fitted with templates from Monte Carlo (MC). The templates are obtained from MC simulations using PYTHIA 8~\cite{Bierlich:2022pfr} with the Monash tune as the event generator, and GEANT4~\cite{Allison:2016lfl,Allison:2006ve, GEANT4:2002zbu} to model the detector response. The templates are constructed, and a linear combination of them is fitted to data as two-dimensional distributions of $DCA_{xy}$ and $DCA_{z}$ in \pT bins. The primary fraction increases from 86\% to 92\% with increasing \pT. Finally, $\lambda$ parameters are determined as a function of \Q, using the information from data on which \pT tracks contribute to each \Q bin. The resulting $\lambda$ parameters are shown in Fig.~\ref{fig:lambdaparam}. 

As described in Section~\ref{sec:analysis}, the p--p--$\Lambda$ correlation function was measured in this work and used to model feed-down contribution to the p--p--p measurement. The \Lam candidates are reconstructed via the weak decay $\Lambda \rightarrow \mathrm{p}\pi^{-}$ (and $\overline{\Lambda}\rightarrow \overline{\mathrm{p}}\pi^{+}$ for $\overline{\Lambda}$ candidates). 
The daughter tracks are selected with criteria similar to those used for primary protons in terms of $|\eta|$ and the number of TPC clusters, but with a looser PID requirement of $n^{\text{PID}}_{\sigma}<5\,(4)$. In addition, the daughter tracks are required to have a DCA to the PV of at least $0.05\,(0.06)$~cm, and the DCA between the two daughters at the secondary vertex must be smaller than $1.5\,(1.2)$~cm. 
The cosine of the pointing angle (CPA), defined as the cosine of the angle between the vector from the PV to the decay vertex and the \Lam candidate momentum, is required to be larger than $0.99\,(0.995)$. Finally, the candidate invariant mass is required to lie between 1.1086 and 1.1226 GeV/$c^2$. The transverse momentum of selected \Lam particles is required to be larger than 0.5 GeV/$c$. The purity of \Lam candidates was estimated from invariant-mass fits, modeling the signal with a double Gaussian and the background with a spline. In addition, template fits to the CPA distribution, similar to the above-described \textit{DCA} fits, were used to extract the primary and secondary \Lam fractions.

An additional selection criterion is applied at the level of two particles, once particle pairs and triplets are formed.
To mitigate track merging effects arising from the finite two-track resolution in the same-event sample, a minimum separation between the two proton tracks is imposed in the $\Delta\eta^*$--$\Delta\varphi^*$ plane for both same- and mixed-event pairs. $\Delta\eta^*$ and $\Delta\varphi^*$, defined as the differences between the pseudorapidities and azimuthal angles of the two particles, respectively, are obtained as averages evaluated at 9 radii in the TPC and corrected for the magnetic-field bending. The default requirement is $\Delta\eta^{*2}+\Delta\varphi^{*2}\ge 0.02^{2}\ (0.022^2)$.

The selection criteria applied to particle candidates constitute the primary source of systematic uncertainties in the measured correlation function. These uncertainties are evaluated by simultaneously varying the selection criteria for protons and $\Lambda$ candidates, as well as for their corresponding antiparticles. The variations are randomly combined into 44 sets, each of which includes at least one variation in the selection requirements. This procedure accounts for
correlations among the systematic uncertainty sources. A given variation set is considered for the evaluation of the associated uncertainty only if the triplet yield changes by less than 40\% relative to the standard selection in the kinematic region $Q_3 < 0.8$~GeV/$c$.

 \subsection{Fit of the p--p correlation function}\label{sec:fits}

The modeled p--p correlation function can be expressed as
\begin{equation}
    C(k^*) = B(k^*)\ [\ \lambda_\mathrm{pp}(k^*)\ C_\mathrm{pp} (k^*) + \lambda_\mathrm{pp_\Lambda}(k^*)\ C_\mathrm{pp_\Lambda}(k^*) + \lambda_\mathrm{others} (k^*)\ C_\mathrm{others}(k^*)\ ].
    \label{eq:Cpp}
\end{equation}
The genuine contribution $C_{\mathrm{pp}}(k^*)$ encodes the final-state interaction of the two protons and is computed by employing the Correlation Analysis Tool using the Schr{\"o}dinger equation (CATS)~\cite{Mihaylov:2018rva}. The CATS framework is capable of calculating the wave function for the particle pair of interest for a given input local potential. In this work, the AV18 and Coulomb potentials are used. The correlation function is calculated by using a Gaussian source function $S(r^*)=1/(4\pi r_\mathrm{0}^{2})^{3/2} \exp[-r^{*2}/(4 r_\mathrm{0}^{2})]$, where $r_\mathrm{0}$ has to be constrained by data. The $C_{\mathrm{pp}_\Lambda}(k^*)$ corresponds to the residual p--$\Lambda$ feed-down. The p–$\Lambda$ interaction is described by chiral EFT calculations at next-to-leading order (NLO19) with a cutoff of 600 MeV~\cite{Haidenbauer:2019boi}, as it provides the best description of all the available p--$\Lambda$ data~\cite{Mihaylov:2023ahn}. The obtained p--$\Lambda$ correlation function is further corrected for the kinematics of the $\Lambda$ decay into protons and pions. Both correlation terms are corrected for experimental momentum resolution effects. All the remaining feed-down contributions and misidentifications provide negligible deviations from unity and are considered as $C_{\mathrm{others}}(k^*) \equiv 1$. The coefficients $\lambda_{ij}(k^*)$ represent the relative weights of the individual components and are determined in a data-driven manner from particle purities and fractions as $\lambda_{ij}(k^*)=P_i(k^*)P_j(k^*)F_i(k^*)F_j(k^*)$. Their averages for $k^* < 200$ MeV/$c$ are $\bar{\lambda}_\mathrm{pp}=0.86$, $\bar{\lambda}_\mathrm{pp_{\Lambda}}=0.05$ and $\bar{\lambda}_\mathrm{others}=0.09$. The factor $B(k^*)$ parametrizes non-femtoscopic background contributions (baseline), and it is modeled by a third-order polynomial without a linear term, $B(k^*) = a\  + b\ k^{*2} + c\ k^{*3}$,
with the parameters $a$, $b$ and $c$ determined simultaneously with $r_{\mathrm{0}}$ from a fit in the range $k^* \in [0,375]$ MeV/$c$.

Uncertainties in the modeling of the correlation function are considered by allowing simultaneous $\pm10\%$ variations of $\lambda_{\mathrm{pp}_\Lambda}(k^*)$ and $\lambda_{\mathrm{others}}(k^*)$, with $\lambda_{\mathrm{pp}}(k^*) = 1 - \lambda_{\mathrm{pp}_\Lambda}(k^*) - \lambda_{\mathrm{others}}(k^*)$, with respect to the nominal values, and changes of the fit range to $k^* \in [0,350]$ and $k^* \in [0,400]$ MeV/$c$. The resulting fit is shown in Fig.~\ref{fig:ppCF}, where the shaded red band represents the $1\sigma$ fit uncertainty obtained with the bootstrap method.

 \subsection{Three-proton correlation function calculation}\label{sec:Norfolk}

\begin{figure}[!ht]
    \centering
    \includegraphics[width=0.65\textwidth]{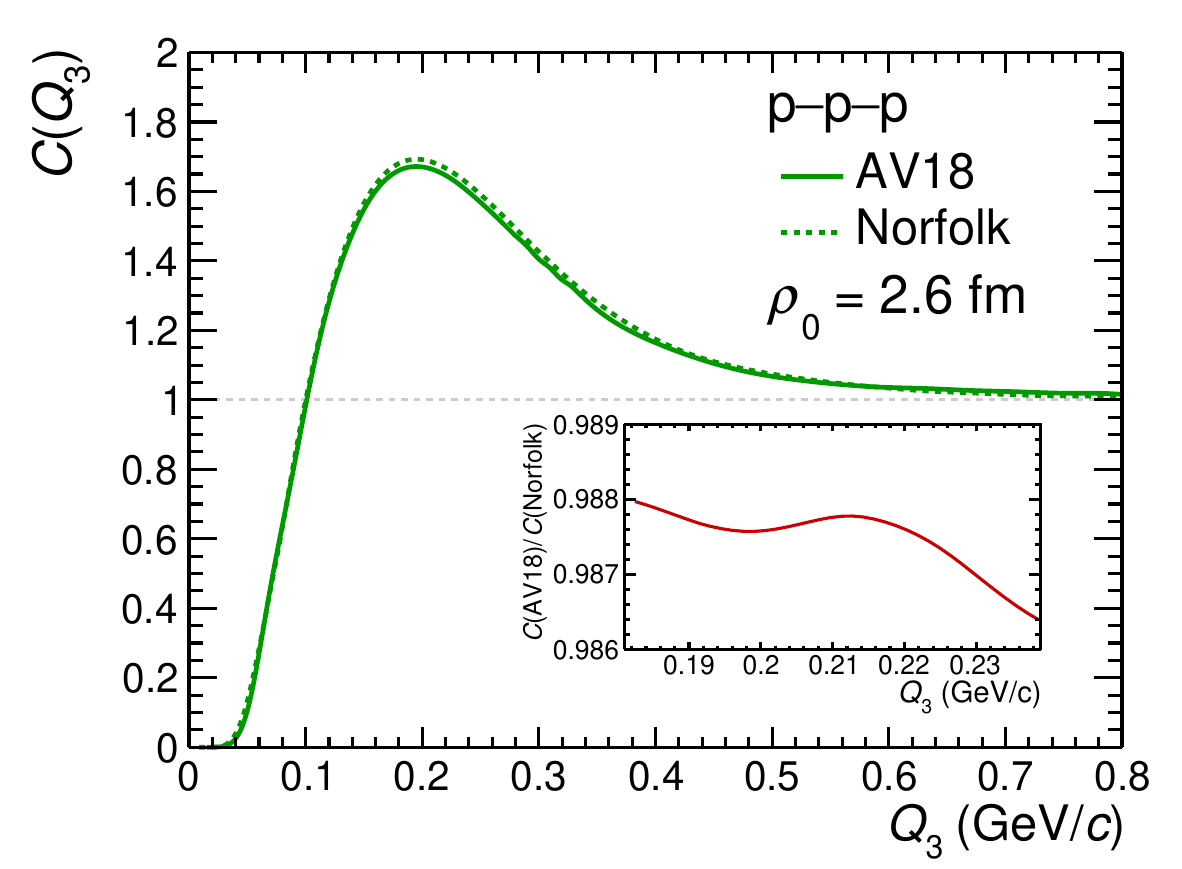}
     \caption{Comparison of the p--p--p correlation function calculated with the AV18 (solid curve) and the Norfolk (dashed curve) potentials for a source size $\rho_0=$ 2.6~fm. The inset shows the ratio of the two correlation functions.}
    \label{fig:norfolk}
\end{figure}

The calculations used for comparison with the measurement employ the AV18 proton--proton potential and Coulomb. While AV18 is among the most widely used phenomenological nucleon--nucleon interactions, more modern chiral potentials have become available in recent decades and can exhibit different short-range behavior. As discussed in Ref.~\cite{Epelbaum:2025aan}, two-body observables are insensitive to the off-shell part of the nucleon--nucleon interaction. Consequently, two potentials that reproduce the two-body observables equally well can predict different three-body observable values. To assess the sensitivity of the calculated p--p--p correlation function to the choice of NN interaction, an additional calculation was performed using a version of the chiral coordinate-space local Norfolk potential introduced in Ref.~\cite{Piarulli:2014bda}. Specifically, the NV2-Ia parametrization from Ref.~\cite{Baroni:2018fdn} was used. As shown in Fig.~\ref{fig:norfolk}, the computed correlation functions obtained with the two potentials are nearly indistinguishable. Only a small difference, below 1.5\%, is observed in the vicinity of the correlation peak, which cannot be resolved with the available precision of the three-proton correlation function measurement. A similar study has been performed in the two-body sector, and the same order of magnitude effect has been observed~\cite{Gobel:2025afq}. Moreover, a slight deviation is observed at higher $Q_3$ values, where the correlation function converges to unity more rapidly when the Norfolk potential is used.

 \subsection{Asymptotic behavior of the correlation function}\label{sec:convergence}

The asymptotic behavior of the correlation function is dictated
by high terms present in the scattering wave function, in turn determined by the large-distance properties of the interactions involved. At the three-body level, as a result of the use of the hyperspherical coordinates, the interactions appear as a series of effective potentials depending on the hyperradius, $\rho$, which are associated asymptotically to specific values of the grand-angular momentum $K$~\cite{Nielsen:2001hbm}.

For large values of $\rho$, one could think that the short-range part of the potential is not active anymore, and the effective potentials, $V_\mathrm{eff}(\rho)$, should then be given simply by the free potential, $V_\mathrm{free}(\rho)$, i.e., by the sum of the 
$K$-dependent three-body centrifugal barrier and the Coulomb potential, $V_\mathrm{Coul}(\rho)$, in the case of dealing with charged particles. However, large $\rho$-values include geometries where two of the particles are close to each other, which implies that, even if $\rho$ is large, the short-range two-body potential is still playing a role. In fact, the asymptotic form of the difference between the effective and free potentials is given by~\cite{Higgins:2020pbe, Higgins:2020avy}
\begin{equation}
    V_\mathrm{eff}(\rho)-V_\mathrm{free}(\rho)=
    V_\mathrm{eff}(\rho)-V_\mathrm{Coul}(\rho)-\frac{\mathchar'26\mkern-9muh^2}{m}\frac{(K+3/2)(K+5/2)}{\rho^2} \stackrel{\rho \rightarrow \infty}{\longrightarrow}
    \frac{\mathchar'26\mkern-9muh^2}{m}\frac{\zeta}{\rho^3} ,
    \label{veff}
\end{equation}
where $m$ is the mass of the proton and $\zeta$ is a constant, having units of length, depending on the angular momentum and parity of the three-body system and on the grand-angular momentum associated with the effective potential. Therefore, the effective potentials do contain a long-range term, whose attractive or repulsive character pushes the correlation function to approach 
the uncorrelated regime (correlation function equal to unity) either from above, if $\zeta<0$, or from below, if $\zeta>0$.   

\begin{figure}[!ht]
    \centering
    \includegraphics[width=0.85\textwidth]{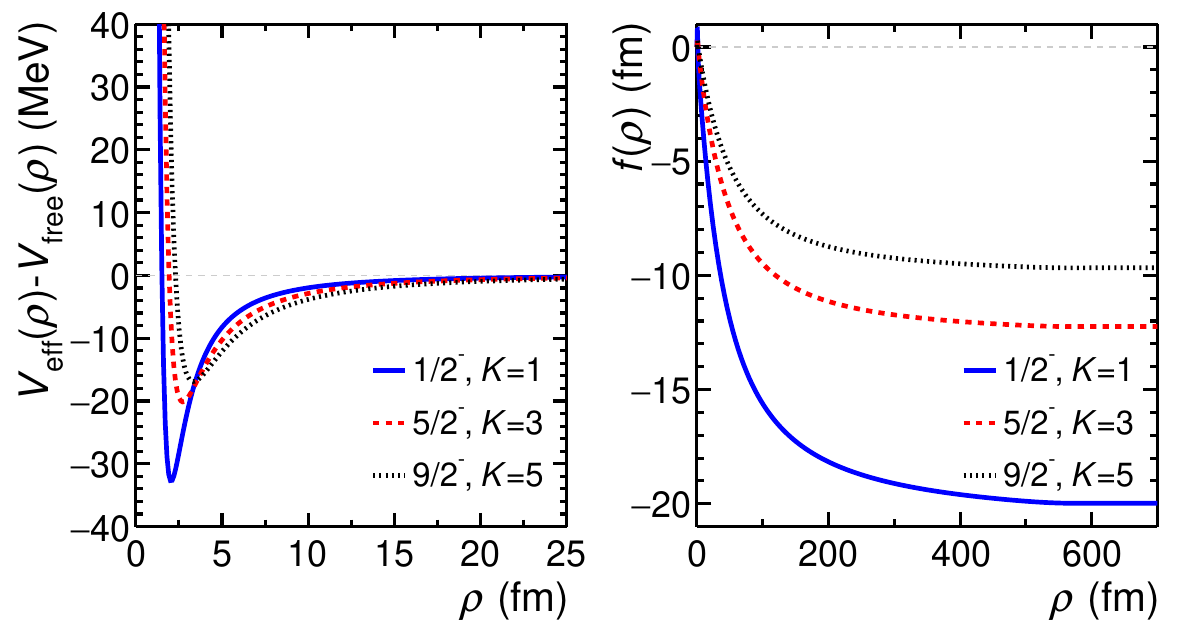}
    \caption{Left: Difference between the effective and free three-body potentials for the $K=1$, $K=3$, and $K=5$ grand-angular momenta in the $1/2^-$, $5/2^-$, and $9/2^-$ three-proton wave functions, respectively. Right: For the same p--p--p states, the function $f(\rho)=(m/\mathchar'26\mkern-9muh^2)(V_\mathrm{eff}(\rho)-V_\mathrm{free}(\rho))\rho^3$ is shown which asymptotically goes to the $\zeta$ constant in Eq.~\ref{veff}. }
    \label{fig:pdppp}
\end{figure}

To illustrate the behavior given in Eq.~\ref{veff}, the left panel of Fig.~\ref{fig:pdppp} shows the function $V_\mathrm{eff}(\rho)-V_\mathrm{free}(\rho)$ for three different potentials in the three-proton system corresponding to the states $J^P = 1/2^-,K=1$ (blue), $J^P = 5/2^-,K=3$ (red), and $J^P=9/2^-,K=5$ (black). Already at small $\rho$-values, the net effect of the additional long-range term contained in the effective potentials is attractive. In the right panel of Fig.~\ref{fig:pdppp}, it is shown, how the function
$f(\rho)=(m/\mathchar'26\mkern-9muh^2)(V_\mathrm{eff}(\rho)-V_\mathrm{free}(\rho))\rho^3$ 
goes asymptotically to the negative constant value $\zeta$. The long-range attraction shown in the left panel of Fig.~\ref{fig:pdppp} is to a large extent responsible for the correlation function being above unity for large $Q_3$ values.

\newpage

%
%

\section{The ALICE Collaboration}
\label{app:collab}
\begin{flushleft} 
\small

D.A.H.~Abdallah\,\orcidlink{0000-0003-4768-2718}\,$^{\rm 137}$, 
I.J.~Abualrob\,\orcidlink{0009-0005-3519-5631}\,$^{\rm 114}$, 
S.~Acharya\,\orcidlink{0000-0002-9213-5329}\,$^{\rm 50}$, 
K.~Agarwal\,\orcidlink{0000-0001-5781-3393}\,$^{\rm II,}$$^{\rm 24}$, 
G.~Aglieri Rinella\,\orcidlink{0000-0002-9611-3696}\,$^{\rm 33}$, 
L.~Aglietta\,\orcidlink{0009-0003-0763-6802}\,$^{\rm 25}$, 
N.~Agrawal\,\orcidlink{0000-0003-0348-9836}\,$^{\rm 26}$, 
Z.~Ahammed\,\orcidlink{0000-0001-5241-7412}\,$^{\rm 135}$, 
S.~Ahmad\,\orcidlink{0000-0003-0497-5705}\,$^{\rm 15}$, 
Z.~Akbar$^{\rm 82}$
% ...

\section*{Affiliation Notes}

$^{\rm I}$ Deceased\\
$^{\rm II}$ Also at: INFN Trieste, Trieste, Italy\\
$^{\rm III}$ Also at: Fondazione Bruno Kessler (FBK), Trento, Italy\\
$^{\rm IV}$ Also at: Czech Technical University in Prague, Prague, Czech Republic\\
$^{\rm V}$ Also at: Dipartimento DET del Politecnico di Torino, Turin, Italy\\
$^{\rm VI}$ Also at: University College of Dublin, Dublin, Ireland\\
$^{\rm VII}$ Also at: Department of Applied Physics, Aligarh Muslim University, Aligarh, India\\
$^{\rm VIII}$ Also at: Institute of Theoretical Physics, University of Wroclaw, Wroclaw, Poland\\
$^{\rm IX}$ Also at: Facultad de Ciencias, Universidad Nacional Aut\'{o}noma de M\'{e}xico, Mexico City, Mexico\\

\section*{Collaboration Institutes}

$^{1}$ A.I. Alikhanyan National Science Laboratory (Yerevan Physics Institute) Foundation, Yerevan, Armenia\\
$^{2}$ AGH University of Krakow, Cracow, Poland\\
$^{3}$ Bogolyubov Institute for Theoretical Physics, National Academy of Sciences of Ukraine, Kyiv, Ukraine\\
$^{4}$ Bose Institute, Department of Physics  and Centre for Astroparticle Physics and Space Science (CAPSS), Kolkata, India\\
$^{5}$ California Polytechnic State University, San Luis Obispo, California, United States\\
$^{6}$ Central China Normal University, Wuhan, China\\
$^{7}$ Centro de Aplicaciones Tecnol\'{o}gicas y Desarrollo Nuclear (CEADEN), Havana, Cuba\\
$^{8}$ Centro de Investigaci\'{o}n y de Estudios Avanzados (CINVESTAV), Mexico City and M\'{e}rida, Mexico\\
$^{9}$ Chicago State University, Chicago, Illinois, United States\\
$^{10}$ China Nuclear Data Center, China Institute of Atomic Energy, Beijing, China\\
$^{11}$ China University of Geosciences, Wuhan, China\\
$^{12}$ Chungbuk National University, Cheongju, Republic of Korea\\
$^{13}$ Comenius University Bratislava, Faculty of Mathematics, Physics and Informatics, Bratislava, Slovak Republic\\
$^{14}$ Creighton University, Omaha, Nebraska, United States\\
$^{15}$ Department of Physics, Aligarh Muslim University, Aligarh, India\\
$^{16}$ Department of Physics, Pusan National University, Pusan, Republic of Korea\\
$^{17}$ Department of Physics, Sejong University, Seoul, Republic of Korea\\
$^{18}$ Department of Physics, University of California, Berkeley, California, United States\\
$^{19}$ Department of Physics, University of Oslo, Oslo, Norway\\
$^{20}$ Department of Physics and Technology, University of Bergen, Bergen, Norway\\
$^{21}$ Dipartimento di Fisica, Universit\`{a} di Pavia, Pavia, Italy\\
$^{22}$ Dipartimento di Fisica, Universit\`{a} di Pisa, Pisa, Italy\\
$^{23}$ Dipartimento di Fisica dell'Universit\`{a} and Sezione INFN, Cagliari, Italy\\
$^{24}$ Dipartimento di Fisica dell'Universit\`{a} and Sezione INFN, Trieste, Italy\\
$^{25}$ Dipartimento di Fisica dell'Universit\`{a} and Sezione INFN, Turin, Italy\\
$^{26}$ Dipartimento di Fisica e Astronomia dell'Universit\`{a} and Sezione INFN, Bologna, Italy\\
$^{27}$ Dipartimento di Fisica e Astronomia dell'Universit\`{a} and Sezione INFN, Catania, Italy\\
$^{28}$ Dipartimento di Fisica e Astronomia dell'Universit\`{a} and Sezione INFN, Padova, Italy\\
$^{29}$ Dipartimento di Fisica `E.R.~Caianiello' dell'Universit\`{a} and Gruppo Collegato INFN, Salerno, Italy\\
$^{30}$ Dipartimento DISAT del Politecnico and Sezione INFN, Turin, Italy\\
$^{31}$ Dipartimento di Scienze MIFT, Universit\`{a} di Messina, Messina, Italy\\
$^{32}$ Dipartimento Interateneo di Fisica `M.~Merlin' and Sezione INFN, Bari, Italy\\
$^{33}$ European Organization for Nuclear Research (CERN), Geneva, Switzerland\\
$^{34}$ Faculty of Electrical Engineering, Mechanical Engineering and Naval Architecture, University of Split, Split, Croatia\\
$^{35}$ Faculty of Nuclear Sciences and Physical Engineering, Czech Technical University in Prague, Prague, Czech Republic\\
$^{36}$ Faculty of Physics, Sofia University, Sofia, Bulgaria\\
$^{37}$ Faculty of Science, P.J.~\v{S}af\'{a}rik University, Ko\v{s}ice, Slovak Republic\\
$^{38}$ Faculty of Technology, Environmental and Social Sciences, Bergen, Norway\\
$^{39}$ Frankfurt Institute for Advanced Studies, Johann Wolfgang Goethe-Universit\"{a}t Frankfurt, Frankfurt, Germany\\
$^{40}$ Fudan University, Shanghai, China\\
$^{41}$ Gauhati University, Department of Physics, Guwahati, India\\
$^{42}$ Helmholtz-Institut f\"{u}r Strahlen- und Kernphysik, Rheinische Friedrich-Wilhelms-Universit\"{a}t Bonn, Bonn, Germany\\
$^{43}$ Helsinki Institute of Physics (HIP), Helsinki, Finland\\
$^{44}$ High Energy Physics Group,  Universidad Aut\'{o}noma de Puebla, Puebla, Mexico\\
$^{45}$ Horia Hulubei National Institute of Physics and Nuclear Engineering, Bucharest, Romania\\
$^{46}$ HUN-REN Wigner Research Centre for Physics, Budapest, Hungary\\
$^{47}$ Indian Institute of Technology Bombay (IIT), Mumbai, India\\
$^{48}$ Indian Institute of Technology Indore, Indore, India\\
$^{49}$ INFN, Laboratori Nazionali di Frascati, Frascati, Italy\\
$^{50}$ INFN, Sezione di Bari, Bari, Italy\\
$^{51}$ INFN, Sezione di Bologna, Bologna, Italy\\
$^{52}$ INFN, Sezione di Cagliari, Cagliari, Italy\\
$^{53}$ INFN, Sezione di Catania, Catania, Italy\\
$^{54}$ INFN, Sezione di Padova, Padova, Italy\\
$^{55}$ INFN, Sezione di Pavia, Pavia, Italy\\
$^{56}$ INFN, Sezione di Pisa, Pisa, Italy\\
$^{57}$ INFN, Sezione di Torino, Turin, Italy\\
$^{58}$ INFN, Sezione di Trieste, Trieste, Italy\\
$^{59}$ Inha University, Incheon, Republic of Korea\\
$^{60}$ Institute for Gravitational and Subatomic Physics (GRASP), Utrecht University/Nikhef, Utrecht, Netherlands\\
$^{61}$ Institute of Experimental Physics, Slovak Academy of Sciences, Ko\v{s}ice, Slovak Republic\\
$^{62}$ Institute of Physics, Homi Bhabha National Institute, Bhubaneswar, India\\
$^{63}$ Institute of Physics of the Czech Academy of Sciences, Prague, Czech Republic\\
$^{64}$ Institute of Space Science (ISS), Bucharest, Romania\\
$^{65}$ Institut f\"{u}r Kernphysik, Johann Wolfgang Goethe-Universit\"{a}t Frankfurt, Frankfurt, Germany\\
$^{66}$ Instituto de Ciencias Nucleares, Universidad Nacional Aut\'{o}noma de M\'{e}xico, Mexico City, Mexico\\
$^{67}$ Instituto de Estructura de la Materia, CSIC, Madrid, Spain\\
$^{68}$ Instituto de F\'{i}sica, Universidade Federal do Rio Grande do Sul (UFRGS), Porto Alegre, Brazil\\
$^{69}$ Instituto de F\'{\i}sica, Universidad Nacional Aut\'{o}noma de M\'{e}xico, Mexico City, Mexico\\
$^{70}$ iThemba LABS, National Research Foundation, Somerset West, South Africa\\
$^{71}$ Jeonbuk National University, Jeonju, Republic of Korea\\
$^{72}$ Korea Institute of Science and Technology Information, Daejeon, Republic of Korea\\
$^{73}$ Laboratoire de Physique Subatomique et de Cosmologie, Universit\'{e} Grenoble-Alpes, CNRS-IN2P3, Grenoble, France\\
$^{74}$ Lawrence Berkeley National Laboratory, Berkeley, California, United States\\
$^{75}$ Lund University Department of Physics, Division of Particle Physics, Lund, Sweden\\
$^{76}$ Marietta Blau Institute, Vienna, Austria\\
$^{77}$ Nagasaki Institute of Applied Science, Nagasaki, Japan\\
$^{78}$ Nara Women{'}s University (NWU), Nara, Japan\\
$^{79}$ National Centre for Nuclear Research, Warsaw, Poland\\
$^{80}$ National Institute of Science Education and Research, Homi Bhabha National Institute, Jatni, India\\
$^{81}$ National Nuclear Research Center, Baku, Azerbaijan\\
$^{82}$ National Research and Innovation Agency - BRIN, Jakarta, Indonesia\\
$^{83}$ Niels Bohr Institute, University of Copenhagen, Copenhagen, Denmark\\
$^{84}$ Nikhef, National institute for subatomic physics, Amsterdam, Netherlands\\
$^{85}$ Nuclear Physics Group, STFC Daresbury Laboratory, Daresbury, United Kingdom\\
$^{86}$ Nuclear Physics Institute of the Czech Academy of Sciences, Husinec-\v{R}e\v{z}, Czech Republic\\
$^{87}$ Oak Ridge National Laboratory, Oak Ridge, Tennessee, United States\\
$^{88}$ Ohio State University, Columbus, Ohio, United States\\
$^{89}$ Physics Department, Panjab University, Chandigarh, India\\
$^{90}$ Physics Department, University of Jammu, Jammu, India\\
$^{91}$ Physics Program and International Institute for Sustainability with Knotted Chiral Meta Matter (WPI-SKCM$^{2}$), Hiroshima University, Hiroshima, Japan\\
$^{92}$ Physikalisches Institut, Eberhard-Karls-Universit\"{a}t T\"{u}bingen, T\"{u}bingen, Germany\\
$^{93}$ Physikalisches Institut, Ruprecht-Karls-Universit\"{a}t Heidelberg, Heidelberg, Germany\\
$^{94}$ Physik Department, Technische Universit\"{a}t M\"{u}nchen, Munich, Germany\\
$^{95}$ Politecnico di Bari and Sezione INFN, Bari, Italy\\
$^{96}$ Research Division and ExtreMe Matter Institute EMMI, GSI Helmholtzzentrum f\"ur Schwerionenforschung GmbH, Darmstadt, Germany\\
$^{97}$ Saga University, Saga, Japan\\
$^{98}$ Saha Institute of Nuclear Physics, Homi Bhabha National Institute, Kolkata, India\\
$^{99}$ School of Physics and Astronomy, University of Birmingham, Birmingham, United Kingdom\\
$^{100}$ Secci\'{o}n F\'{\i}sica, Departamento de Ciencias, Pontificia Universidad Cat\'{o}lica del Per\'{u}, Lima, Peru\\
$^{101}$ SUBATECH, IMT Atlantique, Nantes Universit\'{e}, CNRS-IN2P3, Nantes, France\\
$^{102}$ Sungkyunkwan University, Suwon City, Republic of Korea\\
$^{103}$ Suranaree University of Technology, Nakhon Ratchasima, Thailand\\
$^{104}$ Technical University of Ko\v{s}ice, Ko\v{s}ice, Slovak Republic\\
$^{105}$ The Henryk Niewodniczanski Institute of Nuclear Physics, Polish Academy of Sciences, Cracow, Poland\\
$^{106}$ The University of Texas at Austin, Austin, Texas, United States\\
$^{107}$ Universidad Aut\'{o}noma de Sinaloa, Culiac\'{a}n, Mexico\\
$^{108}$ Universidade de S\~{a}o Paulo (USP), S\~{a}o Paulo, Brazil\\
$^{109}$ Universidade Estadual de Campinas (UNICAMP), Campinas, Brazil\\
$^{110}$ Universidade Federal do ABC, Santo Andre, Brazil\\
$^{111}$ Universitatea Nationala de Stiinta si Tehnologie Politehnica Bucuresti, Bucharest, Romania\\
$^{112}$ University of Cape Town, Cape Town, South Africa\\
$^{113}$ University of Derby, Derby, United Kingdom\\
$^{114}$ University of Houston, Houston, Texas, United States\\
$^{115}$ University of Jyv\"{a}skyl\"{a}, Jyv\"{a}skyl\"{a}, Finland\\
$^{116}$ University of Kansas, Lawrence, Kansas, United States\\
$^{117}$ University of Liverpool, Liverpool, United Kingdom\\
$^{118}$ University of Science and Technology of China, Hefei, China\\
$^{119}$ University of Silesia in Katowice, Katowice, Poland\\
$^{120}$ University of South-Eastern Norway, Kongsberg, Norway\\
$^{121}$ University of Tennessee, Knoxville, Tennessee, United States\\
$^{122}$ University of the Witwatersrand, Johannesburg, South Africa\\
$^{123}$ University of Tokyo, Tokyo, Japan\\
$^{124}$ University of Tsukuba, Tsukuba, Japan\\
$^{125}$ University of Zagreb Faculty of Science, Department of Physics, Zagreb, Croatia\\
$^{126}$ Universit\"{a}t M\"{u}nster, Institut f\"{u}r Kernphysik, M\"{u}nster, Germany\\
$^{127}$ Universit\'{e} Clermont Auvergne, CNRS/IN2P3, LPC, Clermont-Ferrand, France\\
$^{128}$ Universit\'{e} de Lyon, CNRS/IN2P3, Institut de Physique des 2 Infinis de Lyon, Lyon, France\\
$^{129}$ Universit\'{e} de Strasbourg, CNRS, IPHC UMR 7178, F-67000 Strasbourg, France, Strasbourg, France\\
$^{130}$ Universit\'{e} Paris-Saclay, Centre d'Etudes de Saclay (CEA), IRFU, D\'{e}partment de Physique Nucl\'{e}aire (DPhN), Saclay, France\\
$^{131}$ Universit\'{e}  Paris-Saclay, CNRS/IN2P3, IJCLab, Orsay, France\\
$^{132}$ Universit\`{a} degli Studi di Foggia, Foggia, Italy\\
$^{133}$ Universit\`{a} del Piemonte Orientale, Vercelli, Italy\\
$^{134}$ Universit\`{a} di Brescia, Brescia, Italy\\
$^{135}$ Variable Energy Cyclotron Centre, Homi Bhabha National Institute, Kolkata, India\\
$^{136}$ Warsaw University of Technology, Warsaw, Poland\\
$^{137}$ Wayne State University, Detroit, Michigan, United States\\
$^{138}$ Yale University, New Haven, Connecticut, United States\\
$^{139}$ Yildiz Technical University, Istanbul, Turkey\\
$^{140}$ Yonsei University, Seoul, Republic of Korea\\
$^{141}$ Affiliated with an institute formerly covered by a cooperation agreement with CERN\\
$^{142}$ Affiliated with an international laboratory covered by a cooperation agreement with CERN.\\

\end{flushleft} 

\end{document}